\documentclass[a4paper,12pt,fleqn,floatfix]{revtex4-1}
\usepackage{graphicx}
\usepackage{caption, subcaption}
\usepackage{amssymb}
\usepackage{bm}
\usepackage{color}
\usepackage{bookmark}
\usepackage{tabularx}
\usepackage{mathtools}
\usepackage{microtype}
\usepackage{hyperref}
\usepackage{cancel}
\usepackage{relsize}

\usepackage{xcolor}
\newcommand {\be}{\begin{eqnarray}}

\newcommand {\ee}{\end{eqnarray}}

\begin{document}

\title{Negative energy antiferromagnetic instantons forming Cooper-pairing 'glue' and ‘hidden order’ in high-Tc cuprates}

\author{Sergei I. Mukhin}

\affiliation{Theoretical Physics and Quantum Technologies Department, NUST "MISIS", Moscow, Russia}

\begin{abstract}
An emergence of magnetic boson of instantonic nature, that provides  a Cooper-'pairing glue', is considered in the repulsive 'nested' Hubbard model of superconducting cuprates. It is demonstrated, that antiferromagnetic instantons of a spin density wave type may have negative energy due to coupling with Cooper pair condensate. A set of Eliashberg-like equations is derived and solved self-consistently, proving the above suggestion. An instantonic propagator plays the role of the Green function of the pairing 'glue' boson. Simultaneously, the instantons defy condensation of the mean-field SDW order. We had previously demonstrated in analytical form \cite{2,3,4} that periodic chain of instanton-anti-instanton pairs along the axis of Matsubara time has zero scattering cross section for weakly perturbing external probes, like neutrons, etc., thus representing a 'hidden order'. Hence, the two competing orders, superconducting and antiferromagnetic, may coexist (below some Tc) in the form of the mean-field superconducting order coupled to 'hidden' antiferromagnetic one. This new picture is discussed in relation with the mechanism of high temperature superconductivity. 
\end{abstract}

\date{\today}
\maketitle

\section{Introduction}
We present here an idea of instanton-mediated superconductivity using a toy-model Hamiltoninan of electronic system with spin-fermion coupling \cite{chubukov, chubukov1} near the 'nested' Fermi-surface points in momentum space. It proves to be that this model incorporates intrinsic creation of instantons and provides a unified explanation of an emergence of a 'hidden order' state followed by a transition to superconductivity. We start with the spin-fermion model which could be obtained e.g. from a bare on-site repulsive-$U$ Hubbard Hamiltonian with decoupled fermion interaction via auxiliary Hubbard-Stratonovich  field, that allows for collective spin degrees of freedom of the fermi-system. To pay tributes to the symmetries of the assumed short-range spin-ordered state \cite{sachdev_17}, we approximate the field in a form of an assembly of big enough real-space 'spin-bags' of the spin correlation length size, with index $i$ enumerating the bags. Each bag accommodates an antiferromagnetic spin-density wave (SDW) with Matsubara time dependent amplitude ${M}_i(\tau,{\bf{r}})$ and a single wave-vector $\vec{Q}$, and with a 'globally' fluctuating phase $\phi_i = Im \{log M_i(\tau)\}$, that : 

 \begin{eqnarray}
&&H_{HS}=\sum_{q,s}\varepsilon_q{c^{+}_{q,s}}c_{q,s}+\sum_{q,s,i}\left(c^{+}_{q+Q,s}M_i(\tau)s c_{q,s}+H.c.\right)\,\label{HSSDW}\\
&&{M}_i(\tau,{\bf{r}})={M}_i(\tau)e^{i\vec{Q}\vec{r}}+{M_i}^{*}(\tau)e^{-i\vec{Q}\vec{r}},\; M_i(\tau)\equiv |M_i|e^{i\phi_i}
\label{SDWQ}
\end{eqnarray}

\noindent The slow space dependence of a SDW amplitude ${M}_i(\tau,{\bf{r}})$, that delimits the spin-bag volume is not shown explicitly in (\ref{SDWQ}). Since the Hubbard-Stratonovich field must be $\tau$-periodic, the amplitude in (\ref{HSSDW}) obeys periodicity condition:

\begin{eqnarray}
{M}_i\left(\tau+{1}/{T},{\bf{r}}\right)={M}_i(\tau,{\bf{r}}),
\label{MP}
\end{eqnarray}

\noindent where $T$ is temperature. We have absorbed the coupling constant $U$ in the definition of $M$ in the spin-fermion coupling (second) term in (\ref{HSSDW}) . This gives then a renormalized coupling constant  as indicated below in (\ref{L}): $g_{sf}\to g_{sf}U^2$. We shall consider below the case when mean-field SDW order is missing, though $\langle{M}_i(\tau)\rangle$ is 'macroscopic', i.e. proportional to the volume of a 'spin-bag'. This means that an SDW in each spin-bag accommodates instanton-anti-instanton pairs, e.g. considered previously within effective $0+1D$ model \cite{2}. Then, the  following condition is obeyed:

\begin{eqnarray}
\int_0^\beta d\tau \langle{M}_i(\tau)\rangle=0
\label{MP}
\end{eqnarray}

\noindent
Hence, we call such 'invisible SDW' a quantum SDW (QSDW), to emphasize the absence of the mean-field antiferromagnetic order. The bare Lagrangian of this collective bosonic mode is simplified down to the $0+1D$ form:

\begin{eqnarray}
&&L^0 _{AF}  = \frac{1}{2g_{sf}U^2 }\sum_{i}^{N}\left\{ {\dot M_i}^2  + 2\frac{\mu _0 ^2 }{\lambda }{M_i}^2  + {M_i}^4  \right\}\,,\; M_i=\pm |M_i|
\label{L}\\
&&S_{AF}=\int\limits_0^\beta d\tau L^0 _{AF};\;\beta\equiv 1/T
\label{Saf}
\end{eqnarray}

\noindent and real-space antiferromagnetic spin rigidity energy $\propto ({\bf{\nabla}} M_i)^2$ is dropped, being considered as contributing to a 'standard' positive instantonic spin-bag energy shift, while fluctuations of the phase $\phi_i$ are taken into account on the level of 'random phase approximation', i.e. by taking average over $\phi_i$ from $0$ to $2\pi$ in the partition function:

\begin{eqnarray}
Z=Z_fZ_{AF}{\int\;Ad\tau_0{\prod_i\int\cal{D}}\phi_i {\big{\langle}}{\big{\langle}}T_{\tau}\exp\left\{-\int_0^{\beta}\sum_{q,s,i}\left(c^{+}_{q+Q,s}(\tau)M_i(\tau+\tau_0)s c_{q,s}(\tau)+H.c.\right)\right \}  }{\big{\rangle}}_{AF}{\big{\rangle}}_f
\label{ave}
\end{eqnarray}

\noindent Here an interaction representation for the spin-fermion coupling term in Eq. (\ref{HSSDW}) is used \cite{agd}, and Matsubara time $\tau$ ordering procedure is applied to the products of the quantum field-operators, as is indicated with the sign $T_\tau$. The Hibbs averaging  is indicated by the angle brackets $\langle...\rangle$, being performed with the statistical weight provided by the  noninteracting parts of the Hamiltonians/actions of the fermionic and magnetic subsystems, expressed in Eq. (\ref{Saf}) and by the first term in Eq. (\ref{HSSDW}). Integration over $\tau_0$ in the equation (\ref{ave}) arises only  when there is a  zero mode in the $i$-th bag accompanying the instantonic saddle-point solution of the magnetic subsystem described below by equation (\ref{snkt}) and in the text after it. Then, correspondingly, $A$-factor signifies Jacobian used for the integration over the zero mode of the magnetic action $S_{AF}$ \cite{1}: $A\sim \sqrt{S_{AF}^{cl}}$, where $S_{AF}^{cl}=S_{AF}(M_0(\tau))$ is classiacal saddle point value of the instantonic magnetic action. Integration over random phases $\phi_i$ reflects existing symmetry of the spin subsystem on the scale of the 'spin-bag' size, as explained above. We introduce a short-hand notation for the farther convenience:

\begin{eqnarray}
f(\tau)=\sum_{q,s}c^{+}_{q+Q,s}(\tau)s c_{q,s}(\tau)\,;\;f^{\dag}(\tau)=\sum_{q,s}c^{+}_{q,s}(\tau)s c(\tau)_{q+Q,s}
\label{abv}
\end{eqnarray}

\noindent
Now, the time ordering $T_\tau$ permits us to rewrite (\ref{ave}) in the form of series expansion:

\begin{eqnarray}
&&Z=Z_fZ_{AF}\int\;Ad\tau_0\prod_j\int{\cal{D}}\phi_j \sum_{n,m}\dfrac{(-1)^{n+m}}{n!m!}{\big{\langle}}{\big{\langle}}\prod_{i,k=1}^{n}
\prod_{i',k'=1}^{m}\int_0^\beta\,d\tau_k\int_0^\beta\,d{\tau}'_{k'}T_{\tau} f(\tau_k)f^{\dag}({\tau}'_{k'}){\big{\rangle}}_f \times\nonumber\\
&&M_{i}(\tau_k+\tau_0)M^{*}_{i'}({\tau}'_{k'}+\tau_0){\big{\rangle}}_{AF}
\label{aves}
\end{eqnarray}

\noindent Independent averaging over the phases $\phi_i$ of the QSDW in the different 'spin-bags' $i=1,...,N$ gives nonzero result under the conditions: $n=m$ and, simultaneously, couples into "Wick-like" pairwise products the amplitudes $\{M_i\}$: $M_{i}(\tau_k+\tau_0)M_{i'}({\tau}'_{k'}+\tau_0) \delta_{i,i'}$, where $\delta_{ij}$ is Kronecker delta and $M_{i,i'}=\pm |M_{i,i'}|$. Hence, partition function (\ref{ave}) reduces to:

\begin{eqnarray}
&&Z=Z_fZ_{AF}\int\;Ad\tau_0{\big{\langle}}{\big{\langle}}T_{\tau}\exp\left\{-\int_0^{\beta}{\int_0}^{\beta}d{\tau}d{\tau}'\sum_{q,q',s,s',i}D_i(\tau+\tau_0, \tau'+\tau_0)ss' c^{+}_{q+Q,s}(\tau)c_{q,s}(\tau)\times\right.\nonumber\\
&&\left. c^{+}_{q',s'}(\tau')c_{q'+Q,s}(\tau')\right \}{\big{\rangle}}_{AF}{\big{\rangle}}_f;\;D_i(\tau+\tau_0, \tau'+\tau_0)=M_{i}(\tau+\tau_0)M_{i}({\tau}'+\tau_0)
\label{avin}
\end{eqnarray}

\noindent  where all $M_i$ are real. To the lowest order in the four-fermion interaction  term in (\ref{avin}), we substitute retarded interaction $D_i(\tau+\tau_0, {\tau}'+\tau_0)$ by the instantonic propagator ${\cal{D}}(\tau-\tau')$ defined below in Eq. (\ref{Green}).
Hence, we have derived effective retarded interaction between the fermions inside a 'spin-bag', mediated by the fluctuating QSDW.
In what follows, we shall consider instanton-populated 'spin-bags' for the reason explained below. 

\begin{figure}
\begin{center} 
\includegraphics[width=0.75\textwidth]{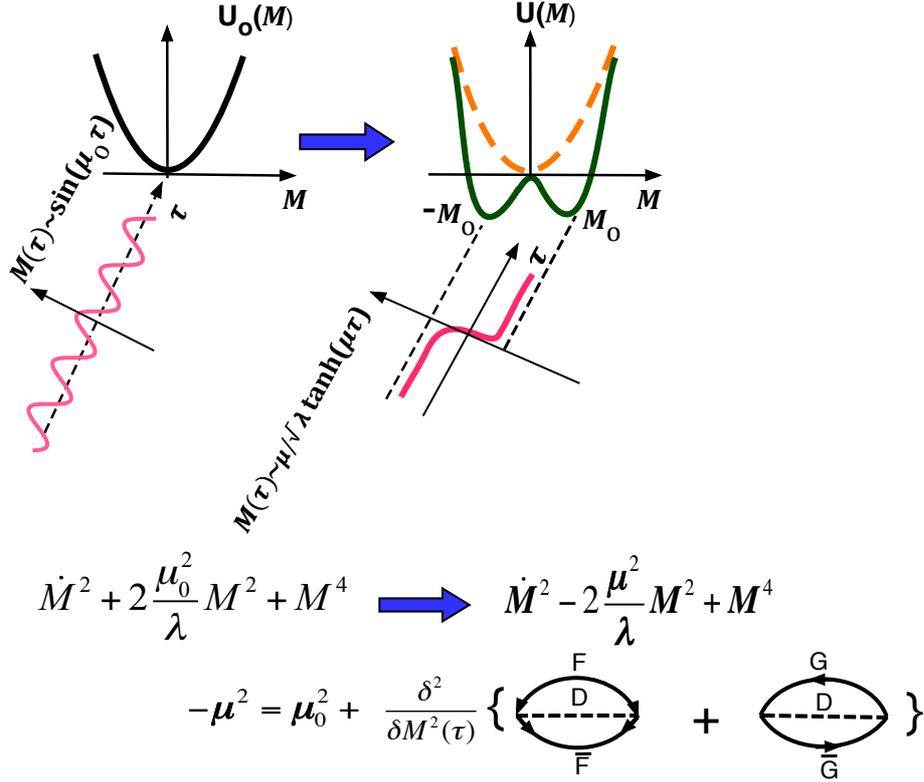}
\end{center}
\caption{ Instanton-mediated Cooper-pairing below T$^{*}$}
\label{1}
\end{figure}

\noindent
Namely, we demonstrate that, under a strong enough spin-fermion coupling in the Hamiltonian (\ref{HSSDW}), a positive bare pre-factor $\mu _0 ^2$ in front of $|M_i|^2$ in (\ref{L}) is renormalised and may become negative: $- \mu ^2$. An intrinsic mechanism of this sign reversal, that happens below a temperature T$^*$, is a first order transition into a phase, that possesses a new saddle point of the Euclidean action of the Fermi-system. The saddle point accommodates  a complex macroscopic fluctuation, that constitutes quantum antiferromagnetic ‘hidden’ order (QSDW) bound to a Cooper-pair condensate inside each 'spin-bag'. We show that as the temperature $T$ is lowered within the temperature interval $T^*<T<T_c$, the energy of this fluctuation crosses zero and becomes negative below T$_c$. This happens due to a growth of the amplitude of the antiferromagnetic QSDW, which is periodically modulated in the imaginary Matsubara time and has zero mean.The latter property makes this QSDW a 'hidden order' \cite{2}. The periodic modulation of the QSDW amplitude along the Matsubara time axis  is  facilitated via sequence of (anti)instantons, an "instantonic crystal", giving rise to instanton-mediated Cooper-pairing 'glue'. The strength of the 'glue' increases as the temperature decreases, and  the energy of the collective fluctuation passes through zero at T$_c$. Below T$_c$ Cooper-pairing fluctuation turns into equilibrium superconducting condensate, and the amplitude of the instantonic modulation of every of the $i=1,..,N$ QSDW saturates and remains finite in the  $T\to 0$ limit.  
A clip-representation of the quint essence of this scenario is presented in Fig. \ref{1}. In the next section we remind derivation \cite{2} of the zero mode instantonic propagator for an {\it{ad hoc}} Lagrangian of the type (\ref{L}), but with sign-changed coefficient $\mu _0 ^2\rightarrow -\mu^2 < 0$,  (\ref{L1}). The 'hidden order' behaviour of the QSDW characterized with this propagator is described. Next, in section III we use the instantonic propagator of section II as a 'glue boson' in the Eliashberg-like scheme of equations, which is derived in the random-phase $\phi_i = Im \{log M_i(\tau)\}$ approximation, and find analytic solution for the temperature Green's functions of the Cooper-paired fermions, using a toy model in Eq. (\ref{HSSDW}), with dispersion $E_q$ possessing "nested" Fermi-surface regions. In section IV a  negative shift of the bare coefficient $\mu_0 ^2$  is calculated explicitly via a second order variational derivative of the free energy decrease, $\Delta\Omega$, due to superconducting fluctuations: $\delta^{2} \Delta\Omega/\delta M^{2}(\tau)$. As a result, an algebraic self-consistency equation for the coefficient $-\mu^2<0$ is obtained and solved. Below a  temperature T$^*$ this coefficient first becomes negative, which manifests transition of the Fermi-system into a state with saddle-point fluctuation described as 'hidden order' inside of each 'spin-bag' accommodating an antiferromagnetic QSDW coupled to superconducting condensate. At strong enough spin-fermion coupling the T$^*$ is greater than T$_c$, giving rise to a 'strange metal' region of the phase diagram of the Fermi-system. Namely, in the interval T$_c<$T$<$T$^*$, as the temperature further decreases below T$^*$, the saddle-point solution splits into two. One of the two saddle-points corresponds to $\mu\propto \mu_0T^*/T$ and has free energy that decreases together with the temperature and at T$_c$ reaches an upper bound of the free energy of the equilibrium superconducting state. Another saddle-point corresponds to $\mu\propto \mu_0T/T^*$ and has free energy that remains higher than the equilibrium free energy value and, hence, remains a fluctuation down to and at T$_c$. Below T$_c$ the  superconducting state coexists with 'hidden' QSDW order, that plays a role of 'pairing glue'. The relevance of the proposed instantonic mechanism of high-temperature superconductivity for cuprates is discussed in the last section V.

\section{Instantonic propagator: Cooper 'pairing glue' and 'hidden order' }

First, we remind our previous derivation \cite{2} of the instantonic propagator, that was obtained using imaginary time-periodic instanton-anti-instanton solution for a Lagrangian of the type (\ref{L}), but with the negative pre-factor in front of $M^2$-term : 

\begin{equation}
\mu _0 ^2  \to  - \mu ^2 ;\;\;L^0 _{AF}  \to L^{eff} _{AF}  = \frac{1}{{2g_{sf} U^2}}\left\{ {\dot M^2  - 2\frac{{\mu ^2 }}{\lambda }M^2  + M^4 } \right\}
\label{L1}
\end{equation}

\noindent
where temperature $T$ and  Matsubara time variable $\tau$ are assumed to be properly renormalized with parameter $\sqrt{\lambda}$: 
$\tilde{\tau}=\tau\sqrt{\lambda/2};\,\tilde{\beta}=\sqrt{\lambda/2}\beta$, and we'll keep track of this in the final answers, avoiding busy formulas in between, compare \cite{1}. Here we also had dropped the spin-bag index $i$ and simplified notations by denoting modulus $|M|$ simply with $M$. It is straightforward to see that saddle-point solution $M_0(\tau)$ of Euclidean action $\tilde{S}_{AF}$ with Lagrangian (\ref{L1}), periodic in the imaginary Matsubara time,  obeys equation for the snoidal Jacobi elliptic function \cite{Witt}. The saddle-point equation is readily derived by equating the variational derivative of the action $\tilde{S}_{AF}$ to zero: 

\begin{eqnarray} 
&&\delta \tilde{S}_{AF}=\delta\int\limits_0^\beta d\tau \frac{1}{{2g_{sf}U^2}}\left\{ {\dot M^2  +(M^2- \frac{{\mu ^2 }}{\lambda })^2-\frac{\mu ^4}{\lambda^2} } \right\}=0; \label{sddl}\\
&&\dot{M}^2=\displaystyle\left({M}^2-\frac{\mu ^2}{\lambda }\right)^2+E\equiv \left({M}^2-\Delta^2\right)\left({M}^2-k^2\Delta^2\right),\label{sn}
\end{eqnarray} 

\noindent where new parameters $\Delta$, $E$ and $k$ are introduced as follows: 

\begin{equation} 
{\Delta ^2(1+k^2) }=2\frac{\mu ^2}{\lambda};\;E=-\frac{\Delta ^4(1-k^2)^2 }{4}
\label{pars}
\end{equation} 

 Indeed, equation (\ref{sn}) has periodic solution expressed via the well known Jacobi snoidal function \cite{Witt}, see Fig. \ref{2}:

\begin{figure}
%\begin{center} 
\includegraphics[width=0.75\textwidth]{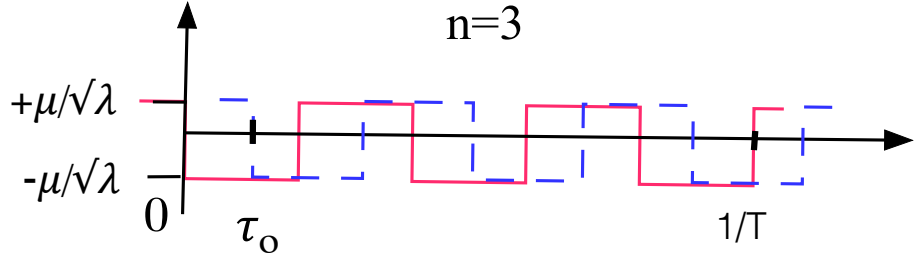}
%\end{center}
\caption{Schematic plot of a periodic saddle-point solution (\ref{snkt}) with the number of instanton-anti-instanton pairs $n=3$. An arbitrary shift $\tau_0$ along the Matsubara axis is indicated with the dashed line, its significance is discussed in the text.}  \label{2}
\end{figure}

\begin{equation} 
M_0(\tau)\equiv k\Delta sn(\Delta\tau;k)\label{snk}.
\end{equation} 

\noindent Here $0\leq k \leq1$ is called elliptic modulus, and Matsubara time periodicity (\ref{MP}) of the saddle-point field $M(\tau)$ imposes conditions:

\begin{equation} 
\Delta=4K(k)nT,\;K(k)=\int\limits_0^1 dx(1-x^2)^{-1/2}(1-k^2x^2)^{-1/2},
\label{del}
\end{equation} 

\noindent where $K(k)$ is elliptic integral of the first kind \cite{Witt}, and $n$ is integer equal to the number of instanton-anti-instanton pairs inside a single period of the Matsubara's time $1/T$, and $T$ is the temperature. Hence, the periodic saddle-point solution is:

\begin{eqnarray}
&&M_0(\tau,\tau_0)\equiv 4kK(k)nT sn(4K(k)nT(\tau+\tau_0),k)=\nonumber\\
&&=k\frac{\mu}{\sqrt{\lambda}}\sqrt{\frac{2}{(1+k^2)}}sn\left(\frac{\mu}{\sqrt{(1+k^2)}}(\tau+\tau_0),k\right).\;\label{snkt}
\end{eqnarray}

\noindent In (\ref{snkt}) a shift $\tau_0$ along the Matsubara axis signifies existence of a zero mode excitation $\propto\partial_{\tau}M_0(\tau)$ causing an arbitrary shift of the saddle-point solution (\ref{snkt}) along the Matsubara time axis without a change of the Euclidean action $\tilde{S}_{AF}(M_0(\tau))$, \cite{1}.
 In passing from the first to last equality in (\ref{snkt}) we had rescaled Matsubara time: $\tau\rightarrow \tau\sqrt{\lambda/2}$, to match notations in \cite{1}. 
 Simultaneously, using the first integral  of the saddle-point differential equation (\ref{sn}) we express the saddle-point action $\tilde{S}_{AF}(M_0(\tau))$ as:

\begin{eqnarray}
&&\tilde{S}_0\equiv\tilde{S}_{AF}(M_0(\tau))=\int\limits_0^\beta d\tau \frac{1}{{2g_{sf}U^2 }}\left\{ {2\dot {M_0}^2 -E-\frac{\mu ^4}{\lambda} }\right\}=\label{act0}\\
&&=\frac{\mu^4}{2g_{sf}U^2T\lambda}\left\{\frac{8n}{3(1+k^2)^{2}K(k)}\left[ (1+k^2)E(k)-(1-k^2)K(k) \right]-\frac{4k^2}{(1+k^2)^2}\right\}.
\label{saf}
\end{eqnarray}

\noindent In the limit $k\to 1$ Jacobi function (\ref{snk}) acquires infinite period $\propto K(k=1)=\infty$ and turns into tangent hyperbolic:

\begin{eqnarray}
&&M_0(\tau;k=1)=\pm\frac{\mu}{\sqrt{\lambda}}\tanh{\left(\frac{\mu}{\sqrt{2}}(\tau+\tau_0)\right)},\;\label{tanh}
\end{eqnarray}

\noindent while $\tilde{S}_0$ becomes $2n$-times the well known single instanton action \cite{1}, but shifted by the mean-field action offset:

\begin{equation}
\tilde{S}_0(k=1)=\frac{1}{2g_{sf}U^2}\left\{2n\left(\frac{2\sqrt{2}\mu^3}{3\lambda}\right)-\frac{\mu^4}{T\lambda}\right\}
\label{sype}
\end{equation}

\noindent The $2n$ factor arises due to imposed Matsubara time periodicity of the Hubbard-Stratonovich field $M(\tau)$, see condition (\ref{MP}), thus leading to an instanton-anti-instanton pairs contribution, with $n$ being the number of such pairs on the interval $[0,1/T]$, the latter being the "thickness" of the Euclidean space slab along the Matsubara time axis. It is important to mention, that combination of  conditions (\ref{pars}) and (\ref{del}) imposes bounds on the independent change of parameters $n$, $k$, and temperature $T$ entering snoidal solution (\ref{snkt}). Namely, to keep $\mu$ finite at $k\rightarrow 1$ one has to assume $nK(k)T\propto \mu=const<\infty$. A choice, that minimises Euclidean action (\ref{sype}), would be to fix $n=1$ and let $TK(k)<\infty$, \cite{1}. We'll return to this later in section IV.

\subsection{Instantonic zero-mode enhancement of the spin-wave 'pairing glue'}
Using instantonic saddle-point solution $M_0(\tau)$ (\ref{snkt}) we define an instantonic propagator:

\begin{equation}
{\cal{D}}(\tau_1-\tau_2,\vec{r}_1-\vec{r}_2)=T\cos(\vec{Q}\cdot (\vec{r}_1-\vec{r}_2))\int_0^{\beta}{ M_0(\tau_1+\tau_0)M_0(\tau_2+\tau_0)}d\tau_0 
\label{Green}
\end{equation}

\noindent The coordinate space dependent pre-factor arises from the nesting wave-vector $Q$ of the QSDW (\ref{SDWQ}). According to the Hubbard Hamiltonian (\ref{HSSDW}), this propagator describes coupling of the fermions to the spin excitations in the saddle-point approximation for ${M}(\tau,{\bf{r}})\to M_{0}(\tau) e^{\pm i\vec{Q}\vec{r}}$, and allows for the zero mode via averaging over $\tau_0$  along the Matsubara time interval $[0,1/T]$. Since we have absorbed the coupling constant $U$ into definition of $M$ in the spin-fermion interaction term in the Hubbard Hamiltonian (\ref{HSSDW}), the spin-density correlator taken in the saddle-point approximation, ${\cal{K}}=\langle T_{\tau}S(\tau_1,\vec{r}_1)S(\tau_2,\vec{r}_2)\rangle$ is related to the propagator  ${\cal{D}}$ in a simple way:

\begin{equation}
{\cal{K}}(\tau_1-\tau_2,\vec{r}_1-\vec{r}_2)=\dfrac{{\cal{D}}(\tau_1-\tau_2,\vec{r}_1-\vec{r}_2)}{U^2}.
\label{K}
\end{equation}

\noindent Now, as it was demonstrated in \cite{2}, propagator ${\cal{D}}$ can be calculated in explicit form from Eqs. (\ref{Green}) and 
 expression $M_0(\tau)$ in Eq. (\ref{snkt}) using Fourier expansion for Jacobi elliptic function $sn$ \cite{Witt}:
 
 \begin{eqnarray}
&& M_0(\tau)=4\pi n T\sum_{m=0}^{\infty}\dfrac{sin(\omega_m\tau)}{\sinh\left(\dfrac{(2m+1) q}{2}\right)};\label{MFour}\;\\
 &&\omega_m=2\pi nT(2m+1);\; q=\pi K(k')/K(k);
\label{jacdefs}
\end{eqnarray}

\noindent  where: $k'^2+k^2=1$. After substitution of expression Eq. (\ref{jacdefs}) into Eq. (\ref{Green}) one finds readily:

\begin{equation}
{\cal{D}}(\tau,\vec{r})= \sum_{m=0}^{\infty}\dfrac{(4\pi nT)^2\cos(\omega_m\tau)\cos(\vec{Q}\cdot\vec{r})}{2\sinh^2\left(\dfrac{(2m+1) q}{2}\right)}.
 \label{weier}
\end{equation}

\noindent Next, the sum in Eq. (\ref{weier}) is expressed via the contour integral \cite{agd} (we dropped Fourier space index $\vec{Q}$ in the argument of the propagator and factor $\cos(\vec{Q}\cdot\vec{r})$) in the r.h.s.of the expression) :

\begin{equation}
{\cal{D}}(\tau)= \dfrac{(2\pi nT)^2}{8\pi i Tn}\int_{C}\dfrac{e^{-2z\tau}\left(1+\tanh\dfrac{z}{2Tn}\right)}{\sinh^2(\dfrac{z q}{2\pi i nT})}dz
 \label{contour}
\end{equation}

\noindent where only the real-space Fourier component with wave-vector $\vec{Q}$ is kept. The integration contour surrounds imaginary axis of $z$, and Matsubara time variable $\tau$ is taken inside the interval: $0<\tau<1/(2nT)$ being the half-period of function $M(\tau)$. Within the latter interval of Matsubara time the integrand in (\ref{contour}) converges fast enough to zero, thus allowing to stretch the contour $C$ along the real axis, leading to equality:

\begin{equation}
{\cal{D}}(\tau)= \dfrac{(2\pi nT)^2 2\pi^2 nT}{q^2}\sum_{s=-\infty}^{+\infty}\dfrac{1}{1+e^{-\frac{z_s}{nT}}}\left[\dfrac {e^{-\frac{z_s}{nT}}} {1+e^{-\frac{z_s}{nT}}}-2\tau\right]e^{-2z_s\tau};\;z_s=\frac{2\pi^2Tns}{q},
 \label{contour1}
\end{equation}

\noindent where summation runs over all integers $s$. In the limit $k\to1$, equivalent to $q\to 0$, see definition in (\ref{jacdefs}), the propagator takes especially simple form, that approaches 'sawtooth' curve along the Matsubara axis, with the period $1/nT$:

\begin{align}
{\cal{D}}(\tau)&= \frac{\pi^2\alpha^2}{8q^2}\left[4nT\tau\left(1-cth\{2\nu nT\tau\}\right)-1+2(1-2nT\tau)cth\{\nu(1-2nT\tau)\}\right];\label{Dd}\\
0\leq\tau&\leq\frac{1}{2nT};\;\alpha^2\equiv{(4\pi nT)^2};\;\nu\equiv\frac{\pi^2}{q}.
\label{DD}
\end{align}

\noindent In the interval ${1}/{2nT}\leq \tau\leq {1}/{nT}$ one finds ${\cal{D(\tau)}}$ using relations:

\begin{equation}
{\cal{D}}(-\tau)={\cal{D}}(\tau); \; {\cal{D}}(\tau)={\cal{D}}\left(\frac{1}{nT}-\tau\right).
 \label{Drels}
\end{equation}

\noindent In Fig. \ref{4} the instantonic propagator (\ref{Dd}) is plotted (blue line), thus manifesting a 'sawtooth' curve.

Finally, approximate expression for the instantonic propagator $D(\tau)$ in the $k\to 1$ limit takes the form below, with relations (\ref{pars}) being used:

\begin{align}
{\cal{D}}(\tau)=\frac{\mu^2}{2\lambda}\begin{cases} [1-4nT\tau];\;0<\tau<{1}/{(2nT)};\\
[4nT\tau-3];\;{1}/{(2nT)}< \tau < {1}/{(nT)}.\end{cases}
\label{Elias51}
\end{align}

\noindent At this point it is convenient to compare the scale of the instantonic propagator found in Eqs. (\ref{Dd}),(\ref{Elias51}): $D(\tau)\propto {\mu^2}/{\lambda}$, with the common spin-wave propagator, see e.g. \cite{chubukov}. For the latter case we use the general recipe of \cite{agd} and find an amplitude of the harmonic oscillator in the vicinity of the local mean-field minima of the Euclidean action (\ref{sddl}) characterised with Lagrangian:  

\begin{equation} 
{L}^{mf}_{AF}=\frac{1}{{2g_{sf}U^2 }}\left\{\dot \delta^2  + \frac{{\mu ^2 }}{\lambda }\delta^2-\frac{\mu ^4}{\lambda^2} \right\},\; {M}_{mf}=\pm\displaystyle\frac{\mu}{\sqrt{\lambda} };\;M={M}_{mf}+\delta. \label{MMF}
\end{equation} 

\noindent From (\ref{MMF}) it is straightforward to check that just opposite to (\ref{Elias51}):

\begin{align}
{\cal{D}}_0(\tau)=-<T_{\tau} \delta(\tau)\delta(0)>\;\propto \frac{\sqrt{\lambda}}{\mu}
\label{DHO}
\end{align} 
 
\noindent Comparison of (\ref{Elias51}) and (\ref{DHO}) indicates, that exchange with instantons in the semiclassical limit: ${\mu^2}/{\lambda}\gg 1$, provides stronger 'pairing glue' than exchange with the spin-waves. Same is true for the spin-waves of the bare Lagranian (\ref{L}), in that case one can use (\ref{DHO}), but exchange $\mu$ for $\mu_0$ in the estimate.

\subsection{Instantonic propagator as 'hidden order'}

Before considering in the next section the role of instantonic exchange in the triggering of superconducting transition at 'high temperature', we first demonstrate why instantonic SDW (i.e. QSDW) is 'hidden order'. 

Namely, it is instructive to use (\ref{Elias51}) and calculate for a particular case of $n=1$ Fourier components of ${\cal{D}}(\tau)$ along the Matsubara axis of bosonic frequencies $\omega_n =2\pi mT$:

\begin{eqnarray}
{\cal{D}}(\omega_m)&=\displaystyle\int_{0}^{1/T}{\cal{D}}(\tau)e^{i\omega_m\tau}d\tau=\propto -\int_{0}^{{1}/{2T}}e^{i\omega_m\tau}\left(\tau-\frac{1}{4T}\right)d\,\tau-\nonumber \\
&-\displaystyle\int_{{1}/{2T}}^{{1}/{T}}e^{i\omega_m\tau}\left(\frac{3}{4T}-\tau\right)d\,\tau=\frac{2}{\omega_m^2}\left(1-(-1)^m\right).
\label{DDF}
\end{eqnarray}

\noindent This calculation demonstrates (proven for the general case in \cite{2}) {\it{a unique property of the propagator  
${\cal{D}}$ to possess  only second order poles}}, i.e. to have zero residues. This comes out from Eq. (\ref{weier}) reflecting the fact that 
 $M_0(\tau)$ in Eq. (\ref{snkt}) is Jacobi's elliptic double periodic function in the complex plane of $\tau$ \cite{Witt}.
\noindent Hence, using the general recipe \cite{agd}, one finds zero cross section $d\sigma(\vec{q}, \omega)$ of the neutron scattering on the instantonic QSDW (\ref{SDWQ})  :

\begin{equation}
d\sigma\sim \dfrac{Im{\cal{D}}^R(\vec{q},\omega)}{U^2(1-\exp{(-\omega/k_BT))}}\equiv 0,
\label{cross}
\end{equation}

\noindent where retarded Green function is obtained by analytic continuation of the propagator (\ref{Green}) from the imaginary Matsubara's axis to the real axis of frequencies, see \cite{2}:  

\begin{eqnarray}
&&{\cal{D}}^R(\omega)\propto -\dfrac{(2\pi Tn)^3}{q^2}\sum_{m=-\infty}^{+\infty}\dfrac{1}{(\omega+2z_m+i\delta)^2}=-\dfrac{\pi Tn}{2sin^2(\tilde{\omega}T_2/4)}\label{rqop}\\
&&z_m=2\pi^2Tnm/q\;;T_2=K(k')/(K(k)nT)\;;\;\tilde{\omega}=\omega+i\delta
\end{eqnarray}

\noindent Hence, we see, indeed, that QSDW (\ref{SDWQ}) has zero scattering cross section in meand field approximation, as it should be since it does not dissipate energy already at finite temperatures. Also the energy transfer $W$ between the external "force" $f(t)$ and the QSDW (\ref{SDWQ}) is strictly zero:

\begin{equation}
W\equiv -i\int_{-\infty}^{+\infty}\dfrac{d\omega}{2\pi}\omega {\cal{D}}^R(\omega)|f(\omega)|^2\equiv 0
\label{invisible}
\end{equation}

\section{Eliashberg equations with instantonic propagator as a Cooper pairing 'glue'}

\begin{figure}
\begin{center} 
\includegraphics[width=0.75\textwidth]{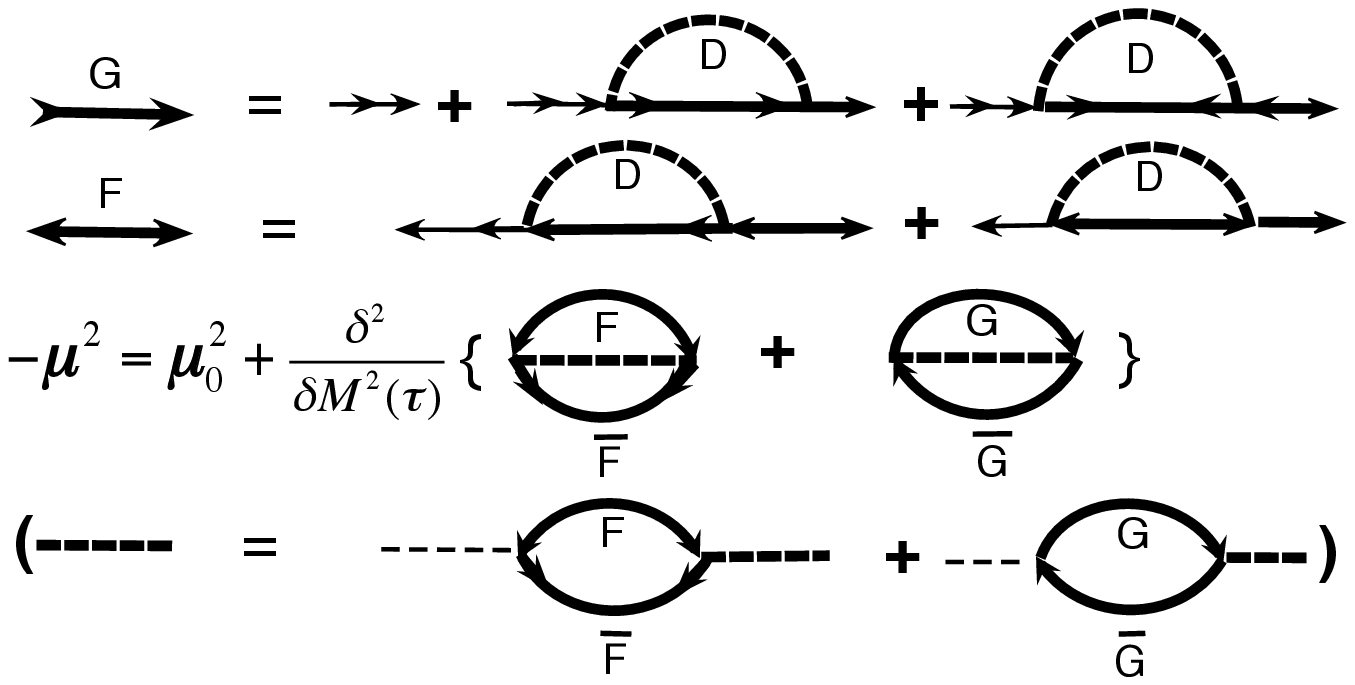}
\end{center}
\caption{The Eliashberg equations, with instantonic spin excitation propagator ${\cal{D}}(\tau)$ of (\ref{Green}) displaying Cooper pairing boson.}
\label{3}
\end{figure}

The Eliashberg equations, with instantonic  propagator ${\cal{D}}(\tau)$ of (\ref{Green}) playing role of spin excitation mode for the Cooper pairing, differ from the common ones \cite{chubukov,elis,emp} by the self-consistency condition applied to the instantonic propagator ${\cal{D}}(\tau)$, as is explained in detail below and symbolically expressed in the last but one line in Fig.\ref{3}. The last line  in Fig.\ref{3} contains the 'common' third equation for the pairing boson propagator and is written in the brackets for comparison, see e.g. \cite{chubukov,emp}. We derive the 'Eliashberg equations' using effective retarded interaction $D_i(\tau+\tau_0, {\tau}'+\tau_0)$ in (\ref{avin}) substituted by the instantonic propagator ${\cal{D}}(\tau-\tau')$ from Eq. (\ref{Green}). Then, the 'usual' integral equations for the self-energy functions $\Sigma _{1p,\sigma }$ and $\Sigma _{2p,\sigma} $ are obtained \cite{elis}.The latter become much simplified under an assumption of the nesting with QSDW's wave-vector $Q$ and a $d$-wave symmetry of the superconducting order parameter in comparison with \cite{elis} (see Appendix for details):

\begin{eqnarray}
&& \Sigma _{1p,\sigma } (\omega ) =\displaystyle \sum\limits_\Omega  {\frac{{{\cal{D}}_Q(\Omega )\left( { - i(\omega  - \Omega ) - \varepsilon _{p - Q}  - \Sigma _{1p - Q,\sigma }^* (\omega  - \Omega )} \right)}}{{| - i(\omega  - \Omega ) + \varepsilon _{p - Q}  + \Sigma _{1,p - Q,\sigma } (\omega  - \Omega )|^2  + |\Sigma _{2p - Q,\sigma } (\omega  - \Omega )|^2 }};} \; \label{sigs1}\\ 
&& \Sigma _{2p,\sigma } (\omega ) = \displaystyle\sum\limits_\Omega  {\frac{{ - {\cal{D}}_Q (\Omega )\Sigma _{2,p - Q,\sigma } (\omega  - \Omega )}}{{| - i(\omega  - \Omega ) + \varepsilon _{p - Q}  + \Sigma _{1p - Q,\sigma } (\omega  - \Omega )|^2  + |\Sigma _{2p - Q,\sigma } (\omega  - \Omega )|^2 }}},
\label{sigs2} 
\end{eqnarray}

\noindent where $\omega=\pi T(2m+1)$ and $\Omega=2\pi T m$, $m=0,\pm 1,..$ are fermionic and bosonic frequencies respectively \cite{agd} . The $d$-wave symmetry of Cooper pairing in combination with 'nesting' conditions for the bare fermionic dispersion leads to the following relations (compare \cite{emp}):

\begin{eqnarray}
&& \varepsilon _{p - Q}  =  - \varepsilon _p\equiv -\varepsilon ;\quad \Sigma _{2p - Q,\sigma }  =  - \Sigma _{2p,\sigma } ;\quad \Sigma _{1p,\sigma }  =  - \Sigma _{1p - Q,\sigma }^* \,;  \label{nesting}\\ 
&& \Sigma _{1p,\sigma } (\omega ) = f(\varepsilon ,\omega ) + is(\varepsilon ,\omega );\:f( - \varepsilon ,\omega ) =  - f(\varepsilon ,\omega );\:s(\varepsilon , - \omega ) =  - s(\varepsilon ,\omega ) .\label{dnes}
 \label{sigma1}
\end{eqnarray}

\noindent In the $k\to 1$ limit the saddle-point action (\ref{saf}) of the spin subsystem  reaches the lowest value, while the instantons acquire a tangent hyperbolic form (\ref{tanh}). Simultaneously, the instantonic propagator $\cal{D}(\tau)$ acquires the sawtooth shape (\ref{Elias51}).  Under these conditions parameter $q$ in (\ref{jacdefs}) becomes small: $q\to 0$, and self-energy function $\Sigma _{1p,\sigma }$ (\ref{dnes}) can be found in algebraic form (see Appendix):

\begin{equation}
\Sigma _{\sigma } (\varepsilon,\omega )=\varepsilon\cdot f+i \omega\cdot s;\; 
\label{fssols}
\end{equation}

\noindent where $f$ and $s$ are slowly dependent on $\omega$ and $\varepsilon$ functions. 

\subsection{Bound states along the axis of Matsubara time}
When conditions (\ref{fssols}) hold, the second Eliashberg equation (\ref{sigs2}) for superconducting self-energy $\Sigma _{2}(\varepsilon ,\omega)$ is transformed into the Schr\"{o}dinger's equation on the Matsubara time axis of coordinates, with instantonic propagator ${\cal{D}}(\tau)$ playing a role of periodic 'potential'. For this purpose we introduce definition of the 'kernel' $K(\tau)\equiv T\sum_\omega K(\omega)e^{-i\omega t}$:

\begin{align}
K(\tau)=T\sum_\omega\dfrac{e^{-i\omega t}}{\omega^2(1-s)^2+\varepsilon^2(1+f)^2+|\Sigma_2|^2}=\frac{{\text{sinh}\left[ {g\left( {\frac{1}{{2T}} - \left| \tau  \right|} \right)} \right]}}{{2g(1 - s)^2 cosh\left( {\frac{g}{{2T}}} \right)}};\label{K} 
\end{align}

\noindent  and: 

\begin{equation}
g^2=\dfrac{\varepsilon^2(1+f)^2+|\Sigma_2|^2}{(1-s)^2}
\label{Ktau0}
\end{equation}

The kernel possesses the following property:

\begin{equation}
\dfrac{\partial^2K(\tau)}{\partial \tau^2}=g^2K(\tau)-\frac{\delta(\tau)}{(1-s)^2}, 
\label{Ktau}
\end{equation} 

\noindent where $\delta(\tau)$ is Dirac Delta function. Above we have approximated self-energy in the denominator of the sum in (\ref{K})  as  $\omega$-independent function of energy $\varepsilon$: $\Sigma _{2}(\varepsilon ,\omega)\to \Sigma _{2}(\varepsilon,0)\equiv \Sigma _{2}$, provided, that $\omega$-dependence of the self-energy $\Sigma _2$ is slow enough and it can be taken at $\omega\approx 0$. 
Using definition  (\ref{K}) of the kernel $K(\tau)$ we introduce new unknown function $\sigma(\varepsilon, \tau)$ instead of $\Sigma _{2,\sigma }(\varepsilon,\omega)$ (the $\varepsilon$ indices are dropped below to simplify notations):

\begin{align}
&\sigma(\omega)\equiv K( \omega) \Sigma _2(\omega),\;\; \sigma \left( \tau  \right) \equiv \int_0^{1/T} {K\left( {\tau  - \tau '} \right)} \Sigma _2 \left( {\tau '} \right)d\tau ',\label{sgma}\\
& \sigma\left( \tau+\frac{1}{T}\right)=-\sigma\left( \tau\right).
\label{sgmap}
\end{align}

\noindent The last antisymmetry condition is due to Fermi-statistics. Then, we rewrite the second Eliashberg equation (\ref{sigs2}) for superconducting self-energy $\Sigma _{2}(\varepsilon ,\omega)$ in the integral form:

\begin{equation}
\sigma \left( {\tau} \right) = \int_0^{{1 \mathord{\left/{\vphantom {1 T}} \right. \kern-\nulldelimiterspace} T}} {K\left( {\tau - \tau '} \right)} {\cal{D}}\left( \tau '  \right)\sigma \left( \tau ' \right)d\tau '.\label{intsgm}
\end{equation}

\noindent Now, using property (\ref{Ktau}) of the kernel $K(\tau)$ and differentiating equation (\ref{intsgm}) twice over $\tau$ we obtain the following Schr\"{o}dinger like equation: 

\begin{equation}
-\sigma ''\left( \tau  \right) - \frac{1}{{(1 - s)^2 }}{\cal{D}}\left( \tau  \right)\sigma \left( \tau  \right) =  - g^2\sigma \left( \tau  \right);\quad {\cal{D}}\left( {\tau  + \frac{1}{{nT}}} \right) = {\cal{D}}\left( \tau  \right),
\label{shrod}
\end{equation}

%%%%%%%%%%%%

\begin{figure}[h!!]
\centering
\begin{subfigure}[h]{0.45\textwidth}
    \centering
    \includegraphics[scale=0.80]{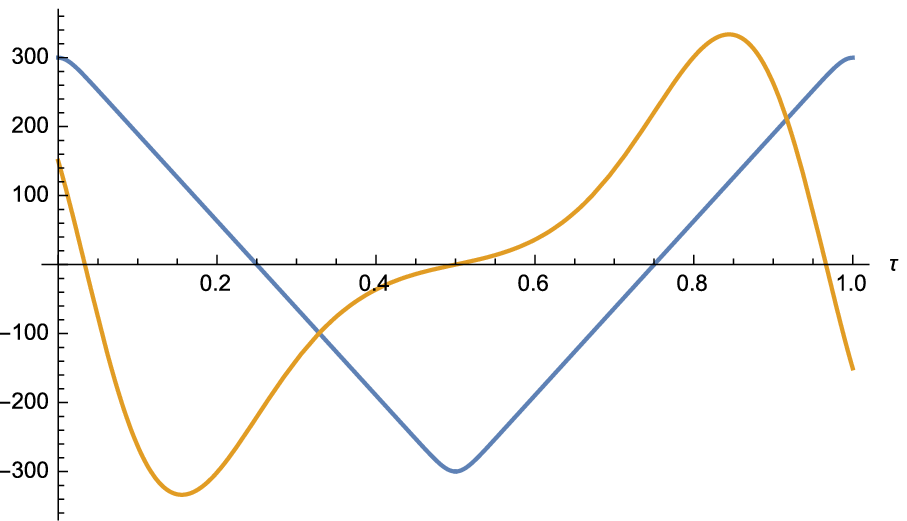}
   \caption{}
    \label{fig:1}
\end{subfigure}%
\begin{subfigure}[h]{0.45\textwidth}
    \centering
    \includegraphics[scale=0.80]{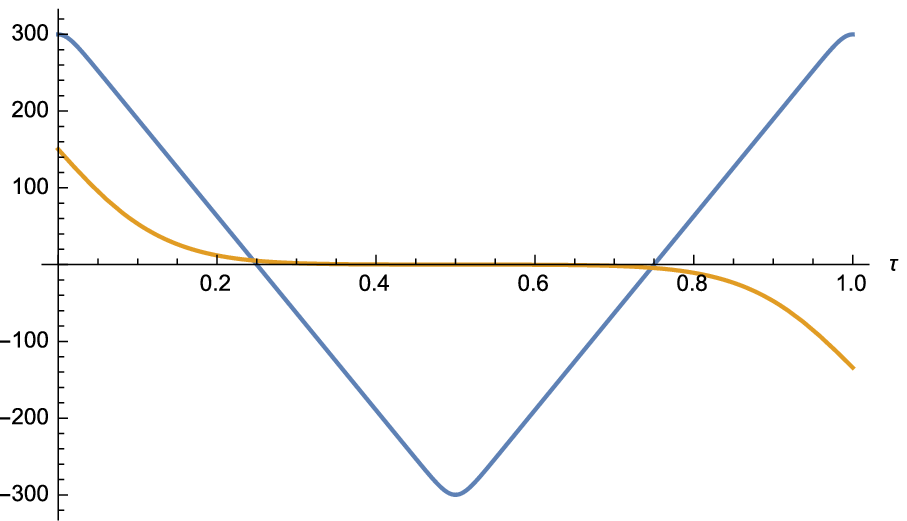}
       \caption{}
    \label{fig:2}
\end{subfigure}
\caption{Effective 'Bloch potential' (blue line) ${\cal{D}}(\tau)/(1-s)^2$ and eigen 'wave function' $\sigma_0(p, \tau)$ (yellow line) corresponding to the following set of parameters: $\mu^2/(2\lambda (1-s)^2)=315.83$, $nT=1$, $T=1$; a) $g_3^2\approx 0.0$; b)$g_1^2=305.34$.}
\label{4}
\end{figure}

\noindent where, indeed, propagator $\cal{D}(\tau)$ plays the role of periodic 'Bloch potential', while unknown function $\sigma(\tau)$ plays the role of the 'wave function', with $-g^2$ being an eigenvalue. According to (\ref{sgmap}) the 'wave function' should posses at least one (odd number of) zero inside the interval $\{0,1/T\}$ of Matsubara slab. Hence, we are looking for the first excited state with eigenvalue $E_1=-g^2_{1}$, which is closest one to the bottom of the 'energy band'. The ground state wave function does not posses zeroes, according to quantum mechanics, and is real and periodic by virtue of the Bloch's theorem, see e.g. \cite{Witt}. Examples of a single zero and a triple zero wave functions, that were calculated numerically, are plotted in Fig.\ref{4}. Now, substituting (\ref{Elias51}) into (\ref{shrod}) we find the following equivalent equation within a single period  $1/nT$ of the 'potential' $\cal{D}(\tau)$:

\begin{equation}
-\sigma ''\left( \tau  \right) + \frac{\mu^24nT}{2\lambda{(1 - s)^2 }}|\tau|\sigma \left( \tau  \right) =  \left(\frac{\mu^2}{2\lambda(1 - s)^2 }- g^2\right)\sigma \left( \tau  \right),\; -\frac{1}{2nT}\leq \tau\leq\frac{1}{2nT}.
\label{shrod1}
\end{equation}

\noindent According to Fig. \ref{fig:2}, the lowest possible eigenvalue $-g^2_{1}$ could be approximated by the minimal value of the sawtooth potential itself, thus, leading to the following solution of the second Eliashberg equation (\ref{sigs2}):

\begin{equation}
{\varepsilon^2(1+f)^2+|\Sigma_2|^2}=g^2_{1}(1-s)^2\approx\frac{\mu^2}{2\lambda}\,.
\label{g1}
\end{equation}

\noindent Hence, nonzero self-energy $\Sigma_2$ exists in the interval of energies around the Fermi-level, $\{-\varepsilon_M,\varepsilon_M\}$:

\begin{equation}
 -\varepsilon_M\leq\varepsilon\leq \varepsilon_M\,;\;\varepsilon_M^2\equiv \{g^2_{1}\frac{(1-s)^2}{(1+f)^2},\,w^2\}|_{min}, 
 \label{e_M}
 \end{equation}

\noindent where $w$ is a width of the energy interval around the bare chemical potential, inside which the nesting condition (\ref{nesting}) holds.
 Other solutions with smaller eigenvalues $-g^2$ do exist as well, see e.g. Fig. \ref{fig:1}, $g^2_{3}= 0;\Sigma_2\equiv 0 $, but they correspond to excited states of Cooper-pairs condensate. 

\section{Instanton driven 'strange metal' and superconducting transitions}

Now we use standard procedure \cite{agd} to calculate free energy change $\Delta\Omega$ per 'spin-bag' due to instanton-mediated superconducting pairing (thus dropping the spin-bag index $i$ introduced in (\ref{HSSDW})) :

\begin{align}
&\Delta\Omega_s=-T\ln\dfrac{Tr\left\{e^{-\int_0^{\beta}H_{int}(\tau)d\tau}{\cal{G}}(0)\right\}}{Tr\left\{{\cal{G}}(0)\right\}}\equiv \Omega_s-\Omega_0;\,{\cal{G}}(0)\equiv e^{-\beta H_0};\,\label{DO} \\
&H_{int}=\left(c^{+}_{q+Q,s}M_0(\tau)s c_{q,s}+H.c.\right)
\label{intinst}
\end{align}

\noindent where $H_0$ is the first term in the sum in (\ref{HSSDW}) respectively. We use the instantonic amplitude $\alpha$ defined in (\ref{DD}), as a formal variable coupling strength in the spin-fermion interaction Hamiltonian $H_{int}$ in (\ref{intinst}) and calculate the free energy derivative:

\begin{align}
\dfrac{\partial\Omega_s}{\partial \alpha}=T{\int_0^{\beta}{\left\langle\dfrac{\partial H_{int}(\tau)}{\partial \alpha}\right\rangle} d\tau}=
-\frac{T}{\alpha}{\int_0^{\beta}\int}_0^{\beta}\left\langle\langle H_{int}(\tau)H_{int}(\tau')\right\rangle\rangle_{\tau_0}d\tau d\tau',
\label{DO1}
\end{align}

\noindent where thermodynamic averaging in (\ref{DO1}) together with an averaging over the 'zero mode' shift $\tau_0$ of the instantons leads to the following relation, see Appendix:

\begin{align}
&\dfrac{\partial\Omega_s}{\partial \alpha}=\frac{T^2}{\alpha}\sum_{\Omega,\omega\,p,\sigma}{\cal{D}}(\Omega)\left\{\dfrac{\overline{\Sigma}_{2,p-Q,\sigma}(\omega)\Sigma_{2p,\sigma}(\omega-\Omega)+{\Sigma}_{2,p-Q,\sigma}(\omega)\overline{\Sigma}_{2p,\sigma}(\omega-\Omega)}{\Phi(\omega)\Phi(\omega-\Omega)}+ \right. \nonumber\\
&\left.+\dfrac{2Re\left[(i\omega+\varepsilon_p - {\Sigma}_{1,p,\sigma}(\omega))(i(\omega-\Omega)+\varepsilon_{p-Q}-{\Sigma}_{1,p-Q,\sigma}(\omega-\Omega))\right]}{\Phi(\omega)\Phi(\omega-\Omega)}\right\}
\label{DO2}
\end{align}

\noindent where:

\begin{equation}
\Phi(\omega)=(i\omega-\varepsilon_p - {\Sigma}_{1,p,\sigma}(\omega))(i\omega+\varepsilon_{p} + {\Sigma}_{1,p,\sigma}^{*}(\omega))-
\Sigma_{2p,\sigma}(\omega)\overline{\Sigma}_{2p,\sigma}(\omega)
\label{Phi}
\end{equation}

\noindent Now, we take into account 'nesting' conditions with vector $\vec{Q}$ expressed in (\ref{nesting}), (\ref{dnes}), and further use definition of 'kernel' $K(\omega)$ in (\ref{K}) in combination with Eliashberg equations (\ref{sigs1}), (\ref{sigs2}). Along this route we finally obtain, after subtraction of the 'normal state' free energy : $\Delta\Omega_s\equiv\Omega_1-\Omega_1({\Sigma}_{2}=0)$, the following expression: 

\begin{equation}
\dfrac{\partial\Delta\Omega_s}{\partial \alpha}=-\frac{2T}{\alpha}\sum_{\omega,p,\sigma}K(\omega)\left|{\Sigma}_{2,p,\sigma}(\omega)\right|^2\approx -\frac{\text{tanh}\left({g_1}/{2T}\right)}{\alpha(1-s)^2g_1}\sum_{p,\sigma}\left|{\Sigma}_{2,p,\sigma}(\omega=0)\right|^2\,,
\label{DOF}
 \end{equation}

\noindent where we had inferred $K(\tau=0)$ from (\ref{K}) and approximated self-energy ${\Sigma}_{2,p,\sigma}(\omega)$ with a frequency independent function of momentum $p$ at $\omega=0$. Now, we use solution (\ref{g1}) for the self-energy ${\Sigma}_{2}$ and pass from summation over momentum $p$ to an integration over energy $\varepsilon=\varepsilon(p)$, simultaneously introducing a bare density of states $\nu_0$ in the vicinity of the Fermi-level. Then, relation (\ref{DOF}) further yields :

\begin{align}
&\frac{\partial\Delta\Omega_s}{\partial \alpha}=-\frac{2 \nu_0}{\alpha{g_1}}\text{tanh}\left(\dfrac{g_1}{2T}\right)\int\limits_0^{\varepsilon_M} d\varepsilon\left[{g_1}^2-\varepsilon^2\dfrac{(1+f)^2}{(1-s)^2}\right]=\nonumber\\
&=-\frac{2\nu_0}{\alpha{g_1}}\text{tanh}\left(\dfrac{g_1}{2T}\right)\left[{g_1}^2\varepsilon_M-\frac{\varepsilon_M^3}{3}\dfrac{(1+f)^2}{(1-s)^2}\right]
\label{DOFg}
\end{align}

\noindent where upper limit of integration $\varepsilon_M$ is defined in (\ref{e_M}).  To proceed, one uses the following relation that follows from Eqs. (\ref{pars}), (\ref{jacdefs}) and (\ref{DD}):

\begin{equation}
g_1^2\equiv\alpha^2 A^2\approx\frac{\mu^2}{2\lambda(1-s)^2}\,;\; A^2=\frac{\pi^2}{8q^2(1-s)^2}.
\label{alpA}
\end{equation}

\noindent Following the well known procedure of calculation of the free energy of an interacting system \cite{agd}, we substitute $\alpha\to x$ in (\ref{DOFg})  and integrate over $x$ from $0$ to $\alpha$, thus, finding $\Delta\Omega_s$:

\begin{equation}
\Delta\Omega_s=\int_0^{\alpha}\dfrac{\partial\Delta\Omega}{\partial x}\,dx=-\frac{4\nu_0}{3}\left|\frac{1-s}{1+f}\right| A^2 \int_0^{\alpha}x\, \text{tanh}\left(\dfrac{xA}{2T}\right) \,dx .
\label{DEL1}
\end{equation}

\noindent  Before we proceed one important observation is in order. The above integration in (\ref{DEL1}) neglects dependence of coefficients $f,s$ on $\alpha$: see Eqs. (\ref{ssmp})-(\ref{smp}). This leads to a simplified result for T$^*$ below, (\ref{tstar}). When allowing for  $\alpha$ dependence of  $f,s$ one finds more involved expression for T$^*$,(\ref{tstars}), that results in Fig. \ref{5}. A detailed derivation will be published in the paper under preparation.  Now, neglecting mentioned above effect, we obtain a simple expression, that depending on ratio $\alpha A/2T$,  has two limits:

\begin{align}
\Delta\Omega_s=-\frac{2\tilde{\nu_0}}{3}\begin{cases} \dfrac{\alpha^3 A^3}{3T}\equiv\dfrac{(g_1^2)^{3/2}}{3T} ;\;{\alpha A}/{2T}\ll1;\\
{\alpha^2 A^2}\equiv g_1^2;\;{\alpha A}/{2T}\gg1.\end{cases};\; \tilde{\nu_0}\equiv \nu_0\left|\frac{1-s}{1+f}\right|.
\label{DELS}
\end{align}

\noindent  Now, using (\ref{DELS}) one is in a position to find self-consistently a phase transition from the bare 'spin-wave' Lagrangian (\ref{L}) to an 'instantonic' Lagrangian (\ref{L1}) (so far assumed {\it{ad hoc}}), which is induced by Cooper pairing fluctuations.  Namely, in the above derivation of (\ref{DELS}) an instantonic pairing 'glue' propagator (\ref{Dd}) was used to evaluate the lowest energy eigenvalue $-g_1^2$ of the 'Schr\"{o}dinger's' equation (\ref{shrod}). Hence, using (\ref{DELS}) we can relate a value of the free energy per 'spin-bag' decrease due to superconducting fluctuations, $\Delta\Omega_s$, to pairing 'glue' amplitude: $g^2_{1}(1-s)^2\approx {\mu^2}/{2\lambda}$, and infer from this a mechanism of (sign)change of the pre-factor: $\mu_0^2\to -\mu^2$ in the Lagrangian (\ref{L1}). A value of parameter $\mu$ has to be determined self-consistently, which is described in the next subsection.

\subsection{Self-consistency equation for instantonic phase formation}
An idea of the following derivation is to cast energy decrease (\ref{DELS}) into a form:

\begin{equation}
\Delta\Omega_s=-{2c^2}T\int\limits_0^\beta d\tau \frac{1}{2g_{sf}U^2}M_0^2\left( \tau  \right),
\label{coef}
\end{equation}

\noindent which then leads to the following expression for effective Euclidean action $\tilde{S}_0$ of the system:

\begin{eqnarray}
&&\tilde{S}_{0} = \dfrac{\Delta\Omega_s}{T}+\int\limits_0^\beta d\tau L^0 _{AF}=\int\limits_0^\beta d\tau\frac{1}{2g_{sf} U^2}\left\{ {\dot M^2  - 2\frac{\mu ^2}{\lambda }M^2  + M^4 } \right\}; \label{SEF}\\
&&\mu ^2 +\mu_0^2=c^2.
\label{cmu1}
\end{eqnarray}

\noindent Here (\ref{cmu1}) follows immediately from definitions (\ref{coef}), (\ref{SEF}) and (\ref{L}).
In order to find coefficient $c^2$ from (\ref{coef}) we calculate variation of the both sides of equality (\ref{coef}) under an infinitesimal variation of the function $M_0\left( \tau  \right)$ at a time instant $\tau$. The variation of the left hand side of (\ref{coef}), $\Delta\Omega_s$, can be found using  well known formula \cite{niv}, that relates variation of e.g. eigenvalue $-g_1^2$ of the Schr\"{o}dinger's equation (\ref{shrod}) to an infinitesimal change of potential ${\cal{D}}\left( \tau  \right)$ at a time instant $\tau$:

\begin{equation}
\delta{g_1^2}_\tau=\frac{1}{(1-s)^2}\delta{\cal{D}}\left( \tau  \right)\sigma_1^*(\tau)\sigma_1(\tau),
\label{var}
\end{equation} 

\noindent where $\sigma_1(\tau)$ is eigenfunction of the Schr\"{o}dinger's equation (\ref{shrod}) corresponding to the eigenvalue $-g_1^2$ , and variation of the potential is derived readily from (\ref{Green}):

\begin{equation}
\delta{\cal{D}}\left( \tau  \right)=\delta M\left( \tau  \right)M_0\left( 2\tau  \right).
\label{varD}
\end{equation}

\noindent Substituting (\ref{varD}) into (\ref{var}) we obtain:

\begin{equation}
\delta{g_1^2}_\tau=\frac{1}{(1-s)^2}\delta M\left( \tau  \right)M_0\left( 2\tau  \right)\sigma_1^*(\tau)\sigma_1(\tau)
\label{varOM}
\end{equation}

\noindent Choosing a zero origin of the Matsubara time interval $\{0,1/T\}$ at $M_0\left( \tau=0  \right)=0$ and taking into account strong localisation of the eigenfunction $\sigma_1(\tau)$ in the vicinities of  the minima of potential ${\cal{D}}\left( \tau  \right)$, see Fig.\ref{fig:2}, we rewrite (\ref{varOM}):

\begin{equation}
\delta{g_1^2}_\tau=\frac{2}{(1-s)^2n}\delta M\left( \tau  \right)M_0\left( \tau  \right),
\label{varOMa}
\end{equation}

\noindent where factor $1/n$ arises due to normalisation of the eigenfunction $\sigma_1(\tau)$ in the $n$ minima of potential ${\cal{D}}\left( \tau  \right)$ possessing period $1/nT$. Using (\ref{varOMa}), it is straightforward to find variation of $\Delta\Omega_s$:

\begin{align}
\delta\left\{\Delta\Omega_s\right\}_{\tau}=-\frac{2\tilde{\nu_0}}{3}\delta M\left( \tau  \right)M_0\left( \tau  \right)\begin{cases}\dfrac{g_1}{T(1-s)^2n} ;\;{g_1}/{2T}\ll1;\\
\dfrac{2}{(1-s)^2n};\;{g_1}/{2T}\gg1.\end{cases}.
\label{varLOMa}
\end{align}

\noindent Simultaneously, variation of the right hand side of (\ref{coef}) is found trivially:

\begin{equation}
\delta\left\{-{2c^2}T\int\limits_0^\beta d\tau \frac{1}{2g_{sf}U^2}M_0^2\left( \tau  \right)\right\}_{\tau}=-\frac{2c^2}{g_{sf}U^2}\delta M\left( \tau  \right)M_0\left( \tau  \right).
\label{varROMa}
\end{equation}

\noindent Now, equating results in (\ref{varLOMa}) and (\ref{varROMa}) and using equation (\ref{cmu}) and known value of $g_1$ from (\ref{g1}) one finds self-consistency equation for the pre-factor $\mu^2$ of the effective instantonic action (\ref{SEF}):

\begin{align}
c^2\equiv \mu^2+\mu_0^2=\frac{\tilde{\nu_0} g_{sf}U^2}{3}\begin{cases}\dfrac{\mu}{T\sqrt{2\lambda}|1-s|^3n} ;\;\dfrac{\mu}{2T\sqrt{2\lambda}|1-s|}\ll1;
\\
\dfrac{2}{(1-s)^2n};\;\dfrac{\mu}{2T\sqrt{2\lambda}|1-s|}\gg1.\end{cases}
\label{cc}
\end{align}

\noindent  Hence, we found that positive bare coefficient  $\mu_0^2$ in Lagrangian (\ref{L}) may turn into a negative coefficient $-\mu^2$ in the effective Lagrangian (\ref{L1}) due to Cooper pair condensate formation,thus, manifesting formation of an 'instantonic phase'. The latter would be manifested by a nonzero constant $g_1^2$, see (\ref{g1}) and Fig.\ref{4}. 

\subsection{'Strange metal' phase below transition temperature T$^*$}
Our strategy is to investigate evolution with temperature of the Euclidean action ${S}_0(T)$ of the system (\ref{sype}), starting from origination of the instantonic phase: ${S}_0(T^*)>0$  (likely called 'strange metal' phase in high-T$_c$ cuprates) till transition to superconducting phase: ${S}_0(T_c)<0$.
We proceed by solving equations (\ref{cc}) simultaneously with Eliashberg equations for the constants $f$ and $s$, defined in (\ref{fssols}), that follow from (\ref{sigs1}), see Appendix.  First, consider the 'high temperatures' interval:  $ g_1/2T\ll1$. Then, the first of the equations (\ref{cc}) constitutes a quadratic equation, and together with  equations for the constants  $f$ and $s$ read:

\begin{eqnarray}
&&\mu^2-\dfrac{2\tilde{G}^2}{T}\mu+\mu_0^2 =0; \; \tilde{G}^2= \dfrac{\tilde{\nu_0} g_{sf}U^2}{6\sqrt{2\lambda}|1-s|^3n}\label{self1};\;\;\tilde{\nu_0}\equiv \nu_0\left|\dfrac{1-s}{1+f}\right|\\
&&s^2-s+G^2=0;\;G^2\equiv\dfrac{\mu^2}{24\lambda n^2T^2};\;\left(\dfrac{g_1}{2T}\right)^2=\dfrac{\mu^2}{8\lambda T^2(1-s)^2}\ll1\label{sG}\\
&&f+1=\dfrac{G^2}{G^2-s^2}.\label{fG}
\label{cmu}
\end{eqnarray}

\noindent  An inequality in (\ref{sG}) hints to smallness of $G^2$ parameter, leading indeed, to a consistent solution:

\begin{eqnarray}
&&s_{\pm}=\frac{1}{2}\left(1\pm\sqrt{1-4{G}^2}\right);\label{ssmp}\\
&&s_{-}\approx G^2;\; f_{-}=\dfrac{s_{-}^2}{G^2-s_{-}^2}\approx {G}^2;\; \label{sfsmp}\\
&&\tilde{\nu_0}\equiv \nu_0\left|\dfrac{1-s}{1+f}\right|\approx \nu_0;\\
&&\mu_{\pm}=\dfrac{\tilde{G}^2}{T}\pm\sqrt{\dfrac{\tilde{G}^4}{T^2}-\mu_0^2};\label{musmp}\\
&&\tilde{G}^2\approx  \dfrac{{\nu_0} g_{sf}U^2}{6\sqrt{2\lambda}n}.
\label{smp}
\end{eqnarray}

 \noindent The choice of "$-$" sign in (\ref{ssmp}) is dictated by consistency with inequality (\ref{sG}). Hence, from (\ref{musmp}) one readily finds a temperature T$^*$, at which transition to an instantonic phase first takes place:
 
 \begin{equation}
 T^*=\dfrac{\tilde{G}^2}{\mu_0}\left|_{n=1}\approx \dfrac{{\nu_0} g_{sf}U^2}{6\sqrt{2\lambda}\mu_0}\right.
 \label{tstar}
\end{equation}

\noindent In relation with remark made after Eq. (\ref{DEL1}), an account of $s,f$ dependence on $\alpha$ leads to a more involved relation (derivation is pending in the paper under preparation):

 \begin{equation}
 {T^*}^2= \dfrac{({\nu_0} g_{sf}U^2)^2}{36(\mu_0\sqrt{\lambda})^2}\left[1\pm\sqrt{1-\dfrac{12\mu_0^4}{({\nu_0} g_{sf}U^2)^2}}\right],
 \label{tstars}
\end{equation}

\noindent where the upper sign brunch leads to result (\ref{tstar}), while the lower sign brunch leads to a saturation of T$^*$ at $\mu_0/(\sqrt{6\lambda})$ in the large limit of dimensionless coupling constant $g=\sqrt{{\nu_0} g_{sf}U^2}/\mu_0$. The two brunches of the instantonic amplitude  $\mu_{\pm}$ originate at T$^*$ according to (\ref{musmp}), and split in the temperatures interval $T^*>T>T_c$ while starting from the common initial value $\mu_{\pm}(T=T^*)=\mu_0$:

\begin{eqnarray}
&&\mu_{+}(T)\approx 2\dfrac{\tilde{G}^2}{T};\;  \;\label{mus}
\\
&&\mu_{-}(T)\approx T\dfrac{\mu_0^2}{2\tilde{G}^2};\;{T^*>>T>T_c}
\label{mut}
\end{eqnarray}

\noindent where both expressions are given in the 'low temperature' limit: $T_c <<T<<T^*$. In order for these solutions to exist the following condition must hold:

\begin{equation}
G(T^*)=\dfrac{\mu(T^*)}{2\sqrt{6\lambda}{T^*}}\equiv\dfrac{\mu_0^2}{\sqrt{3}{\nu_0} g_{sf}U^2}<1/2.
\label{condstr}
\end{equation}

\noindent Thus, temperature dependences  of the Euclidean action ${S}_0(T)$ of the system (\ref{sype}) corresponding to the two instantonic brunches $\mu_{\pm}(T)$ differ. While brunch $\mu_{+}(T)$  finally leads to a condensation of Cooper pairs in superconducting state at T$_c$, the other brunch $\mu_{-}(T)$ remains a (macroscopic) fluctuation mode, that gradually softens (${S}_0(\mu_{-}(T))\propto T^3$) as the temperature decreases. 

\subsection{Superconducting transition inside the instantonic phase: T$_c$} 
 
Consider now an expression for the effective Euclidean action ${S}_0(T)$ of the system (\ref{sype}) with normal metal Euclidean action being subtracted, see (\ref{SEF}). It is obvious, that transition from instantonic phase to superconducting thermal equilibrium state is manifested by ${S}_0(T)$ becoming negative. Hence, equation that defines superconducting transition temperature  T$_c$ is just: 

\begin{equation}
\tilde{S}_0(T_c)=\frac{1}{2g_{sf}U^2}\left\{2n\left(\frac{2\sqrt{2}\mu(T_c)^3}{3\lambda}\right)-\frac{\mu(T_c)^4}{T_c\lambda}\right\}=0;\;n=1.
\label{eupm}
\end{equation}

\noindent It is straightforward to infer from (\ref{eupm}) and definition (\ref{alpA}) that:

\begin{equation}
\frac{\mu(T_c)}{T_c\sqrt{\lambda}}=2\left(\frac{2\sqrt{2}}{3\sqrt{\lambda}}\right);\;\dfrac{g_1}{2T_c}=\dfrac{\mu(T_c)}{2T_c\sqrt{2\lambda}|1-s|}\equiv \dfrac{2}{3 \sqrt{\lambda}|1-s|}\gg1.
\label{neq}
\end{equation}

\noindent Hence, in the vicinity of T$_c$ one has to use the second of equations (\ref{cc}) and also equations for the constants  $f$ and $s$, that are valid in the limit:  $\dfrac{g_1}{2T_c}\gg1$ (see Appendix):

\begin{eqnarray}
&&\mu^2+\mu_0^2=\frac{2\tilde{\nu_0} g_{sf}U^2}{3(1-s)^2};\label{self2}\\
&&s^2-s-G_1^2=0;\;G_1^2\equiv\dfrac{2\mu^2}{\lambda g_1^2};\;\label{sG2}\\
&&f=\dfrac{s^2}{G_1^2-s^2}.\label{fG2}\\
&&{g_1^2}=\dfrac{\mu^2}{2\lambda(1-s)^2}\;\label{g1sc}.
\label{cmusu}
\end{eqnarray}

\noindent Next, one substitutes (\ref{sG2}) into (\ref{fG2}), and also (\ref{g1sc}) into (\ref{sG2}), leading after a simple algebra to the following relations:

\begin{align}
f=-s;
\begin{cases}s=1; \\
s=\frac{4}{3}.\end{cases}\label{sc}
\end{align}

\noindent Then, a choice consistent with inequality (\ref{g1sc}) and finiteness of the instantonic amplitude in (\ref{self2}) would be: 

\begin{equation}
s=-f=\frac{4}{3}.
\label{fssc}
\end{equation}

\noindent Finally, substituting (\ref{fssc}) into (\ref{self2}) one finds T$_c$ from (\ref{neq}):

\begin{equation}
T_c=\frac{3}{4\sqrt{2}}\left(6\nu_0g_{sf}U^2-\mu_0^2\right)^{1/2}.
\label{tc}
\end{equation}

\begin{figure}
\begin{center} 
\includegraphics[width=0.75\textwidth]{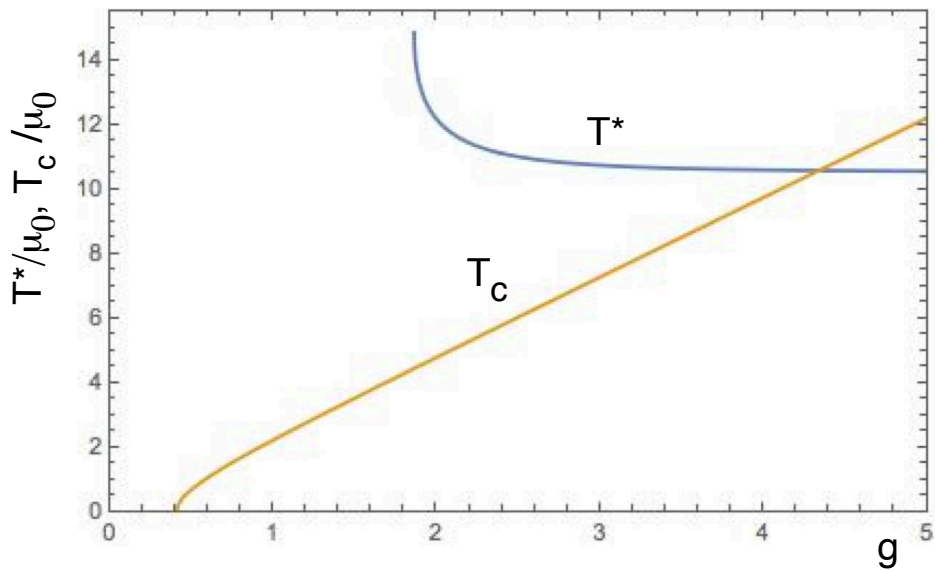}
\end{center}
\caption{ Analytically evaluated schematic plot of the instanton mediated Cooper-pairing  T$^{*}$ and superconducting T$_c$ dependences on effective instanton-fermion dimensionless coupling strength  $g=(\sqrt{\nu_0 g_{sf}}U)/\mu_0$.}
\label{5}
\end{figure}

\noindent It is interesting to observe, that a necessary condition for existence of solution for  T$_c$ follows from (\ref{tc}):

\begin{equation}
\dfrac{\nu_0g_{sf}U^2}{\mu_0^2}>\dfrac{1}{6},
\label{condtc}
\end{equation}

\noindent and is less restrictive than condition for existence of T$^*$ solution in (\ref{tstars}) : ${\nu_0} g_{sf}U^2/\mu_0^2>\sqrt{12}$. Thus, there may exist an interval of intermediate coupling strength: $1/ 6 < \nu_0 g_{sf} U^2/{\mu_0^2} < \sqrt{12}$, in which T$_c$ is not preceded by T$^*$, i.e. 'strange metal' phase is absent above the superconducting dome. This feature is indeed present in the phase diagram of high-T$_c$ cuprates in the 'underdoped' regime \cite{sato,talon,zhang}. Both transition temperatures are found from Eliashberg like system of equations, but with spin wave instantonic propagator playing role of pairing boson.  Fig. \ref{5} contains plots of the analytically evaluated T$^*$, (\ref{tstar}), and T$_c$, (\ref{tc}), dependences on effective instanton-fermion dimensionless coupling strength  $g=(\sqrt{\nu_0 g_{sf}}U)/\mu_0$, that surprisingly resembles phase diagram in the temperature-doping coordinates, see e.g. \cite{sato,talon,zhang}. To get the second part of the superconducting T$_c$ dome in the 'overdoped' region of high-T$_c$ cuprates an assumption should be made on the dependences on doping of e.g. bare density of 'nested' fermionic  states $\nu_0$, (\ref{DOFg}), and related cut-off energy $\varepsilon_M$, (\ref{e_M}). Simultaneously, a numerical self-consistent solution of the 'Eliashberg equations' (\ref{sigs1}), (\ref{sigs2}) and (\ref{varROMa}) in the whole interval of coupling $g$ should be made.   
Finally, we mention that  transition temperatures T$^*$ and T$_c$ derived above depend on the powers of  $\sqrt{\nu_0 g_{sf}}U$, rather than on $U\exp\{-1/(\nu_0 g_{sf})\}$ typical  for a weak-coupling BCS theory, compare e.g.{\cite{chubukov1}}.

\section{Conclusions}
\label{sec: fin}
To summarise, an instantonic mechanism of high temperature superconductivity is proposed as part of a wider picture. Namely, it is demonstrated that in principle, an instantonic quantum nematic  'crystal' can emerge as a hidden order that self-consistently provides pairing glue for Cooper pair condensate. Depending on the strength of effective spin-fermion coupling, a temperature of nematic phase transition, T$^*$,  either precedes superconducting transition temperature T$_c$, or ceases to exist, with instantonic quantum nematic emerging together with the superconducting Cooper pair condensate. Quantumness of emergent nematic state is provided by periodic in Matsubara time instantonic modulation of the amplitude of  'hidden' SDW  order. A more detailed calculation of the 'spin-bag' instanton-anti-instanton configuration in 2+1D Euclidean space is in progress and will be presented elsewhere.    
  
\section*{Acknowledgement}
The author acknowledges useful discussions with Jan Zaanen, Konstantin Efetov, Serguey Brazovskii and Andrey Chubukov, as well as partial support of this work by the Russian Ministry of science and education Increase Competitiveness Program of NUST MISiS (No. K2-2017-085).

\appendix

\section{Self-energy parts and Dyson equations}
\label{Elias_s}
\noindent Let $G$ and $F$ be normal and anomalous fermionic Green's functions respectively, where ${\cal{D}}$ is  instantonic Green's function (\ref{Green}), compare\cite{2}. Then, normal and anomalous self-energy parts of the fermionic Green's functions, $\Sigma_1$ and $\Sigma_2$ respectively, take the form:

\begin{align}
\Sigma_{1p,\sigma}(\omega) &= T \sum_{\Omega} D_{Q}(\Omega) G_{p-Q,\sigma}(\omega-\Omega) \label{Sigma1}\\
\Sigma_{2p,\sigma}(\omega) &= T \sum_{\Omega} D_{Q}(\Omega) F_{p-Q,\sigma}(\omega-\Omega) \label{Sigma2}\\
\Sigma_{1,-p,\overline{\sigma}}(-\omega) &= T \sum_{\Omega} D_{Q}(\Omega) G_{-p+Q,\overline{\sigma}}(-\omega+\Omega)\label{oSigma1} \\
\overline{\Sigma}_{2p,\sigma}(\omega) &= T \sum_{\Omega} D_{Q}(\Omega) \overline{F}_{p-Q,\sigma}(\omega-\Omega) \label{oSigma2}\\
\Sigma_{1,p-Q,\sigma}(\omega) &= T\sum_{\Omega} D_{Q}(\Omega) G_{p,\sigma}(\omega-\Omega) \label{Sigma1p-Q}\\
\Sigma_{2,p-Q,\sigma}(\omega) &= T\sum_{\Omega} D_{Q}(\Omega) F_{p,\sigma}(\omega-\Omega) \label{Sigma2p-Q}\\
\Sigma_{1,-p+Q,\overline{\sigma}}(-\omega) &= T\sum_{\Omega} D_{Q}(\Omega) G_{-p,\overline{\sigma}}(-\omega+\Omega)\label{oSigma1p-Q} \\
\overline{\Sigma}_{2,p-Q,\sigma}(\omega) &= T\sum_{\Omega} D_{Q}(\Omega) \overline{F}_{p,\sigma}(\omega-\Omega) \label{oSigma2p-Q}.
\end{align}

\noindent Now, having the list above, one derives a closed set of  the Dyson equations, that will be solved in algebraic form with respect to the yet unknown Green functions expressed via the self-energies to be found from the Eliashberg equations derived below.

%\section{The Dyson equations}
%\label{Elias_d} 
\noindent  A set of Dyson equations based on the Hamiltonian (\ref{HSSDW}) is as follows:

\begin{align}
&(i\omega - \varepsilon_p) G_{p,\sigma}(\omega)= 1 + \Sigma_{1p,\sigma}(\omega) G_{p,\sigma}(\omega) + \Sigma_{2p,\sigma} \overline{F}_{p,\sigma}(\omega); \label{E1}\\
&(i\omega + \varepsilon_{p}) \overline{F}_{p,\sigma}(\omega)=  -\Sigma_{1,-p,\overline{\sigma}}(-\omega) \overline{F}_{p,\sigma}(\omega) +
\overline{\Sigma}_{2p,\sigma} G_{p,\sigma}(\omega); \;\label{E2}\\
&(-i\omega - \varepsilon_p) G_{-p,\overline{\sigma}}(-\omega) = 1 + \Sigma_{1,-p,\overline{\sigma}}(-\omega) G_{-p,\overline{\sigma}}(-\omega) + \overline{\Sigma}_{2p,\sigma} F_{p,\sigma}(\omega);\; \label{E3}\\
&(-i\omega + \varepsilon_p) F_{p,\sigma}(\omega) =  -\Sigma_{1p,\sigma}(\omega) F_{p,\sigma}(\omega) + \Sigma_{2p,\sigma} G_{-p,\overline{\sigma}}(-\omega);\; \label{E4}\\
&(i\omega - \varepsilon_{p-Q}) G_{p-Q,\sigma}(\omega) = 1 + \Sigma_{1,p-Q,\sigma}(\omega) G_{p-Q,\sigma}(\omega) + \Sigma_{2,p-Q,\sigma} \overline{F}_{p-Q,\sigma}(\omega);\; \label{E5}\\
&(i\omega + \varepsilon_{p-Q}) \overline{F}_{p-Q,\sigma}(\omega) =  -\Sigma_{1,-p+Q,\overline{\sigma}}(-\omega) \overline{F}_{p-Q,\sigma}(\omega) +
\overline{\Sigma}_{2,p-Q,\sigma} G_{p-Q,\sigma}(\omega);\; \label{E6}\\
&(-i\omega - \varepsilon_{p-Q}) G_{-p+Q,\overline{\sigma}}(-\omega) = 1 + \Sigma_{1,-p+Q,\overline{\sigma}}(-\omega) G_{-p+Q,\overline{\sigma}}(-\omega) +\nonumber\\
& \overline{\Sigma}_{2,p-Q,\sigma} F_{p-Q,\sigma}(\omega);\; \label{E7}\\
&(-i\omega + \varepsilon_{p-Q}) F_{p-Q,\sigma}(\omega) =  -\Sigma_{1,p-Q,\sigma}(\omega) F_{p-Q,\sigma}(\omega) + \Sigma_{2,p-Q,\sigma} G_{-p+Q,\overline{\sigma}}(-\omega).\;\label{E8}
\end{align}

\noindent Solving the algebraic system of equations (\ref{E1}) - (\ref{E8}) for $G$'s and ${F}$'s we find (introducing shorthand notation: $\overline{\Sigma}_{1p,\sigma}(\omega)\equiv\Sigma_{1,-p,\overline{\sigma}}(-\omega)$):

\begin{align}
&G_{p,\sigma}(\omega) = \frac{ -i \omega - \varepsilon_p -\overline{\Sigma}_{1p}}
{(i\omega+\varepsilon_p+\overline{\Sigma}_{1p})(-i\omega+\varepsilon_p+ \Sigma_{1p}) + \Sigma_{2p}\overline{\Sigma}_{2p}}; \label{Gp}\\
&\overline{F}_{p,\sigma}(\omega) = \frac{-\overline{\Sigma}_{2p,\sigma}}
{(i\omega+\varepsilon_p+\Sigma_{1,-p,\overline{\sigma}}(-\omega))(-i\omega+\varepsilon_p+ \Sigma_{1p,\sigma}(\omega)) + \Sigma_{2p,\sigma}\overline{\Sigma}_{2p,\sigma}};\label{oFp} \\
&G_{-p,\overline{\sigma}}(-\omega) = \frac{i \omega - \varepsilon_p -\Sigma_{1p,\sigma}(\omega)}
{(-i\omega+\varepsilon_p+\Sigma_{1p,\sigma}(\omega))(i\omega+\varepsilon_p+ \Sigma_{1,-p,\overline{\sigma}}(-\omega)) + \Sigma_{2p,\sigma}\overline{\Sigma}_{2p,\sigma}}
\label{oGp} \\
&F_{p,\sigma}(\omega) = \frac{-\Sigma_{2p,\sigma}}
{(-i\omega+\varepsilon_p+\Sigma_{1p,\sigma}(\omega))(i\omega+\varepsilon_p+ \Sigma_{1,-p,\overline{\sigma}}(-\omega)) + \Sigma_{2p,\sigma}\overline{\Sigma}_{2p,\sigma}};\;\label{Fp}\\
&G_{p-Q,\sigma}(\omega) = \frac{- i \omega - \varepsilon_{p-Q} -\Sigma_{1,-p+Q,\overline{\sigma}}(-\omega)}
{(i\omega+\varepsilon_{p-Q}+\Sigma_{1,-p+Q,\overline{\sigma}}(-\omega))(-i\omega+\varepsilon_{p-Q}+ \Sigma_{1,p-Q,\sigma}(\omega)) + \Sigma_{2,p-Q,\sigma}\overline{\Sigma}_{2,p-Q,\sigma}}; \label{Gp-Q}\\
&\overline{F}_{p-Q,\sigma}(\omega) = \frac{-\overline{\Sigma}_{2,p-Q,\sigma}}
{(i\omega+\varepsilon_{p-Q}+\Sigma_{1,-p+Q,\overline{\sigma}}(-\omega))(-i\omega+\varepsilon_{p-Q}+ \Sigma_{1,p-Q,\sigma}(\omega)) + \Sigma_{2,p-Q,\sigma}\overline{\Sigma}_{2,p-Q,\sigma}};\;\label{oFp-Q}\\
&G_{-p+Q,\overline{\sigma}}(-\omega) = \frac{i \omega - \varepsilon_{p-Q} -\Sigma_{1,p-Q,\sigma}}
{(-i\omega+\varepsilon_{p-Q}+\Sigma_{1,p-Q,\sigma}(\omega))(i\omega+\varepsilon_{p-Q}+ \Sigma_{1,-p+Q,\overline{\sigma}}(-\omega)) + \Sigma_{2,p-Q,\sigma}\overline{\Sigma}_{2,p-Q,\sigma}};\;\label{oGp-Q} \\
&F_{p-Q,\sigma}(\omega) = \frac{-\Sigma_{2,p-Q,\sigma}}
{(-i\omega+\varepsilon_{p-Q}+\Sigma_{1,p-Q,\sigma}(\omega))(i\omega+\varepsilon_{p-Q}+ \Sigma_{1,-p+Q,\overline{\sigma}}(-\omega)) + \Sigma_{2,p-Q,\sigma}\overline{\Sigma}_{2,p-Q,\sigma}}.\;\label{Fp-Q}
\end{align}

\noindent Now, using the above expressions for the Green's functions we provide derivation, that leads from Eq. (\ref{DO1}) to Eq. (\ref{DO2}):

\begin{align}
\dfrac{\partial\Omega_s}{\partial \alpha}=
-\frac{T}{\alpha}{\int_0^{\beta}\int}_0^{\beta}\left\langle\langle H_{int}(\tau)H_{int}(\tau')\right\rangle\rangle_{\tau_0}d\tau d\tau'=\frac{T^2}{\alpha}\sum_{\Omega,\omega\,p,\sigma}{\cal{D}}(\Omega)\mathcal{G}_{p,\sigma}(\omega)\mathcal{G}_{p,\sigma}(\omega-\Omega);
\label{DOFF}
\end{align} 

\noindent where a product of the generalised Greeen's functions reads:

\begin{align}
&\mathcal{G}_{p,\sigma}(\omega)\mathcal{G}_{p,\sigma}(\omega-\Omega)=G_{p,\sigma}(\omega)G_{p-Q,\sigma}(\omega-\Omega)+G_{-p,\sigma}(\omega-\Omega)G_{-p+Q,\sigma}(\omega)+\nonumber\\
&\overline{F}_{p,\sigma}(\omega)F_{p-Q,\sigma}(\omega-\Omega)+F_{p,\overline{\sigma}}(\omega)\overline{F}_{p-Q,\overline{\sigma}}(\omega-\Omega)
\label{PROD}
\end{align}

\noindent Now, substituting into (\ref{PROD}) the above expressions for the Green's functions (\ref{Gp})-(\ref{Fp-Q}) and taking into account relations (\ref{symms}) derived in section \ref{Elias_sf} below, one obtains Eq. (\ref{DO2}) in the main text.

\section{Eliashberg equations}
\label{Elias_sf}
\noindent Now, substituting into equations (\ref{Sigma1})-(\ref{oSigma2p-Q}) relations (\ref{Gp})-(\ref{Fp-Q}), and allowing for a relation: $\overline{\Sigma}_{2,p-Q,\sigma}(\omega-\Omega)=\Sigma^*_{2,p-Q,\sigma}(\omega-\Omega)$, to be checked below {\it{a posteriori}}, we obtain eight coupled Eliashberg equations:

\begin{align}
&\Sigma_{1p,\sigma}(\omega) = T\sum_{\Omega} {D_{Q}(\Omega) \left(- i (\omega-\Omega) - \varepsilon_{p-Q} -{\Sigma}_{1,-p+Q,\overline{\sigma}}(-\omega+\Omega)\right)}\left[ (i(\omega-\Omega)+ \varepsilon_{p-Q}+ \right. \nonumber\\
&\left. {\Sigma}_{1,-p+Q,\overline{\sigma}}(-\omega+\Omega))(-i(\omega-\Omega)+\varepsilon_{p-Q}+ \Sigma_{1,p-Q,\sigma}(\omega-\Omega)) + |{\Sigma_{2,p-Q,\sigma}(\omega-\Omega)}|^2 \right]^{-1}; \label{Eliash8_Sigma1p}\\
&\Sigma_{2p,\sigma}(\omega) = -T\sum_{\Omega} D_{Q}(\Omega) \Sigma_{2,p-Q,\sigma}(\omega-\Omega)\left[
(-i(\omega-\Omega)+\varepsilon_{p-Q}+\right. \nonumber\\
&\left. \Sigma_{1,p-Q,\sigma}(\omega-\Omega))(i(\omega-\Omega)+\varepsilon_{p-Q}+ {\Sigma}_{1,-p+Q,\overline{\sigma}}(-\omega+\Omega)) + |\Sigma_{2,p-Q,\sigma}(\omega-\Omega)|^2 \right]^{-1}; \label{Eliash8_Sigma2p}\\
& \Sigma_{1,-p,\overline{\sigma}}(-\omega)  = T\sum_{\Omega}D_{Q}(\Omega) \left( i (\omega-\Omega) - \varepsilon_{p-Q} -\Sigma_{1,p-Q,\sigma}(\omega-\Omega) \right)\left[(-i(\omega-\Omega)+\varepsilon_{p-Q}+\right.\nonumber\\
&\left.\Sigma_{1,p-Q,\sigma}(\omega-\Omega))(i(\omega-\Omega)+\varepsilon_{p-Q}+ \Sigma_{1,-p+Q,\overline{\sigma}}(-\omega+\Omega) ) +| \Sigma_{2,p-Q,\sigma}(\omega-\Omega)|^2 \right]^{-1}
\label{Eliash8_oSigma1p};\\
& \overline{\Sigma}_{2p,\sigma}(\omega) =-T\sum_{\Omega}
D_{Q}(\Omega) \overline{\Sigma}_{2,p-Q,\sigma}(\omega-\Omega)
\left[(i(\omega-\Omega)+\varepsilon_{p-Q}+\right.\nonumber\\
&\left.\Sigma_{1,-p+Q,\overline{\sigma}}(-\omega+\Omega))(-i(\omega-\Omega)+\varepsilon_{p-Q}+  \Sigma_{1,p-Q,\sigma}(\omega-\Omega)) + |\Sigma_{2,p-Q,\sigma}(\omega-\Omega)|^2\right]^{-1} 
\label{Eliash8_oSigma2p};\\
& \Sigma_{1,p-Q,\sigma}(\omega) = T\sum_{\Omega}D_{Q}(\Omega) \left( -i (\omega-\Omega) - \varepsilon_p -\Sigma_{1,-p,\overline{\sigma}}(-\omega+\Omega) \right)\left[(i(\omega-\Omega)+\varepsilon_p+\right.\nonumber\\
&\left. \Sigma_{1,-p,\overline{\sigma}}(-\omega+\Omega))(-i(\omega-\Omega)+\varepsilon_p+ \Sigma_{1p,\sigma}(\omega-\Omega)) +|\Sigma_{2p,\sigma}(\omega-\Omega)|^{2}\right]^{-1} 
\label{Eliash8_Sigma1p-Q};\\
& \Sigma_{2,p-Q,\sigma}(\omega) = -T\sum_{\Omega}D_{Q}(\Omega)\Sigma_{2p,\sigma}(\omega-\Omega)\left[
(-i(\omega-\Omega)+\varepsilon_p+\right.\nonumber\\
&\left. \Sigma_{1p,\sigma}(\omega-\Omega))(i(\omega-\Omega)+\varepsilon_p+ \Sigma_{1,-p,\overline{\sigma}}(-\omega+\Omega)) + |\Sigma_{2p,\sigma}(\omega-\Omega)|^{2}\right]^{-1} \label{Eliash8_Sigma2p-Q};\\
& \Sigma_{1,-p+Q,\overline{\sigma}}(-\omega) = T\sum_{\Omega}D_{Q}(\Omega) \left( i (\omega-\Omega) - \varepsilon_p -\Sigma_{1p,\sigma}(\omega-\Omega) \right)\left[(-i(\omega-\Omega)+\varepsilon_p+\right.\nonumber\\
&\left. \Sigma_{1p,\sigma}(\omega-\Omega))(i(\omega-\Omega)+\varepsilon_p+ \Sigma_{1,-p,\overline{\sigma}}(-\omega+\Omega)) + |\Sigma_{2p,\sigma}(\omega-\Omega)|^{2}\right]^{-1} \label{Eliash8_oSigma1p-Q};\\
&\overline{\Sigma}_{2,p-Q,\sigma}(\omega) = -T\sum_{\Omega}D_{Q}(\Omega)\overline{\Sigma}_{2p,\sigma}(\omega-\Omega)\left[
(i(\omega-\Omega)+\varepsilon_p+\right.\nonumber\\
&\left. \Sigma_{1,-p,\overline{\sigma}}(-\omega+\Omega))(-i(\omega-\Omega)+\varepsilon_p+ \Sigma_{1p,\sigma}(\omega-\Omega)) + |\Sigma_{2p,\sigma}(\omega-\Omega)|^{2}\right]^{-1} \label{Eliash8_oSigma2p-Q}.
\end{align}

\noindent
It is easy to check that above equations admit the following relations:

\begin{align}
&\overline{\Sigma}_{2,p,\sigma}(\omega) = {\Sigma}^{*}_{2,p,\sigma}(\omega);\;
\overline{\Sigma}_{1p,\sigma}(\omega)\equiv\Sigma_{1,-p,\overline{\sigma}}(-\omega) = \Sigma^{*}_{1p,\sigma}(\omega);\;\Sigma_{1,p-Q,{\sigma}}(\omega) = -\Sigma^{*}_{1p,\sigma}(\omega).
\label{symms}
\end{align}

\noindent 
In this case we have only four independent Eliashberg equations (\ref{Eliash8_Sigma1p}), (\ref{Eliash8_Sigma2p}), (\ref{Eliash8_Sigma1p-Q}), and (\ref{Eliash8_Sigma2p-Q}), that acquire compact form:

\begin{align}
\Sigma_{1p,\sigma}(\omega) &= T\sum_{\Omega} 
\frac{D_{Q}(\Omega) \left(- i (\omega-\Omega) - \varepsilon_{p-Q} -\Sigma^*_{1,p-Q,\sigma}(\omega-\Omega)\right)}
{ |-i(\omega-\Omega)+\varepsilon_{p-Q}+ \Sigma_{1,p-Q,\sigma}(\omega-\Omega)|^2 + |\Sigma_{2,p-Q,\sigma}(\omega-\Omega)|^2}; \label{Si1}\\
\Sigma_{2p,\sigma}(\omega) &= T\sum_{\Omega}  
\frac{-D_{Q}(\Omega) \Sigma_{2,p-Q,\sigma}(\omega-\Omega)}
{|-i(\omega-\Omega)+\varepsilon_{p-Q}+\Sigma_{1,p-Q,\sigma}(\omega-\Omega)|^2 + |\Sigma_{2,p-Q,\sigma}(\omega-\Omega)|^2};\label{Si2} \\
\Sigma_{1,p-Q,\sigma}(\omega) &= T\sum_{\Omega}
\frac{D_{Q}(\Omega) \left( -i (\omega-\Omega) - \varepsilon_p -\Sigma^*_{1,p,\sigma}(\omega-\Omega) \right)}
{|-i(\omega-\Omega)+\varepsilon_p+ \Sigma_{1p,\sigma}(\omega-\Omega)|^2 + |\Sigma_{2p,\sigma}(\omega-\Omega)|^2} ;\label{Si1m}\\
\Sigma_{2,p-Q,\sigma}(\omega) &= T \sum_{\Omega}
\frac{-D_{Q}(\Omega)\Sigma_{2p,\sigma}(\omega-\Omega)}
{|-i(\omega-\Omega)+\varepsilon_p+\Sigma_{1p,\sigma}(\omega-\Omega)|^2  + |\Sigma_{2p,\sigma}(\omega-\Omega)|^2}\label{Si2m}.
\end{align}

\noindent Now, it is straightforward to check that combined 'nesting' and d-wave symmetry relations (\ref{nesting}) reduce four equations (\ref{Si1})-(\ref{Si2m}) to the two equations in the main text: (\ref{sigs1}) and (\ref{sigs2}). Solutions for $\Sigma_{1p,\sigma}$ and $\Sigma_{2p,\sigma}$ of the latter couple of equations  might be sought for in the form (\ref{fssols}) and (\ref{shrod}) respectively. Combining (\ref{symms}) with (\ref{fssols}) and applying these relations to equation (\ref{Si1}), we find equations for the 'constants' $f$ and $s$ assumed to be slow functions (approximately independent of) $\omega$ and $\varepsilon$ respectively:  

\begin{align}
\varepsilon f-i\omega s\equiv \Sigma^*_{1p,\sigma}(\varepsilon,\omega) =T\sum_{\Omega} 
\frac{D_{Q}(\Omega) \left(i (\omega-\Omega) +\varepsilon +\varepsilon f-is(\omega-\Omega)\right)}
{ |i(\omega-\Omega)+\varepsilon+\varepsilon f-is (\omega-\Omega)|^2 + |\Sigma_{2}|^2}. \label{fseq}
\end{align} 

\noindent Equation (\ref{fseq}) splits into two algebraic equations for the constants $f$ and $s$, and after taking into account expression for the instantonic propagator $D_{Q}(\Omega)$, (\ref{weier}), one finds:

\begin{align}
&f=-\dfrac{\alpha^2(1+f)}{8nT(1-s)^2}\left\{\dfrac{4nT\pi^2}{q^2[(i\omega)^2-g^2]}+\dfrac{\text{tanh}\dfrac{i\omega+g}{4nT}}{2g\text{sin}^2\left[\dfrac{(i\omega+g)q}{4\pi nT}\right]}-\dfrac{\text{tanh}\dfrac{i\omega-g}{4nT}}{2g\text{sin}^2\left[\dfrac{(i\omega-g)q}{4\pi nT}\right]}+\right.\nonumber\\
&\left. \dfrac{\alpha^2}{q^2}\sum_{k=1}^{\infty}\left\{\dfrac{1}{4nT\text{cosh}^2\dfrac{z_k}{4nT}}\left[\dfrac{1}{(z_k-i\omega-g)(z_k-i\omega+g)}+\dfrac{1}
{(z_k+i\omega+g)(z_k+i\omega-g)}\right]-\right.\right. \nonumber\\
&\left.\left. 2{\text{tanh}}\dfrac{z_k}{4nT}\left[ \dfrac{z_k-i\omega}{(z_k-i\omega-g)^2(z_k-i\omega+g)^2}+
\dfrac{z_k+i\omega}{(z_k+i\omega+g)^2(z_k+i\omega-g)^2} \right]\right\}\right\};
\label{fgen}\\
&s\cdot\omega=\dfrac{\alpha^2}{16\pi nT(1-s)} \left\{ -\dfrac{2\pi^2\omega 4\pi nT}{q^2(\omega^2+g^2)} + \pi i\left[ \dfrac{\text{tanh}\dfrac{i\omega+g}{4nT}}
{\text{sin}^2\left[\dfrac{(i\omega+g)q}{4\pi nT}\right]} +\dfrac{ \text{tanh}\dfrac{i\omega-g}{4nT}}{ \text{sin}^2\left[\dfrac{(i\omega-g)q}{4\pi nT}\right]} \right]-\right. \nonumber\\
&\left. 2\pi i \dfrac{\alpha^2}{q^2}\sum_{k=1}^{\infty}  \left\{\dfrac{1}{4nT\text{cosh}^2\dfrac{z_k}{4nT}}\left[\dfrac{i\omega-z_k}{(z_k-i\omega-g)(z_k-i\omega+g)}+
\dfrac{i\omega+z_k}{(z_k+i\omega+g)(z_k+i\omega-g)}\right]- \right.\right.\nonumber\\
&\left.\left. {\text{tanh}}\dfrac{z_k}{4nT}\left[ \dfrac{1}{(z_k+i\omega+g)(z_k+i\omega-g)}- \dfrac{1}
{(z_k-i\omega-g)(z_k-i\omega+g)}+\right.\right.\right.\nonumber\\
&\left.\left.\left. \dfrac{2(z_k+i\omega)^2}{(z_k+i\omega+g)^2(z_k+i\omega-g)^2}-\dfrac{2(z_k-i\omega)^2}{(z_k-i\omega-g)^2(z_k-i\omega+g)^2} \right]\right\}\right\}.
\label{sgen}
\end{align}

\noindent Where the following notations defined previously in equations (\ref{jacdefs}), (\ref{contour1}) and (\ref{DD}), (\ref{g1}) are as follows:

\begin{equation}
\alpha\equiv{(4\pi nT)};\;q=\pi K(k')/K(k);\; z_k=\frac{2\pi^2Tnk}{q};\;
{\varepsilon^2(1+f)^2+|\Sigma_2|^2}=g^2(1-s)^2
\label{notes}
\end{equation}

\noindent where $K(k)$ is elliptic integral of the first kind \cite{Witt}, and we neglected $\omega$-dependence of $\Sigma_2$, as explained in the main text after equation (\ref{Ktau}). Next, we consider limit $k\to1$, equivalent to $q\to 0$, since it corresponds to the least energy per instanton, as explained in the text after equation (\ref{sype}). Two limits could be treated in analytic form: i) $g\ll nT$, and $g\gg nT$. We start with the general case $n\geq 1$, but ultimately will consider $n=1$,  as explained after equation (\ref{sype}) in the main text. 
\subsubsection{High temperatures limit: $g\ll nT$}
Expanding hyperbolic tangents in small parameter $g/nT$ in the numerators in (\ref{fgen}) and (\ref{sgen}) as well as trigonometric sine functions in small parameter $q$ in denominators, one finds the main contributions $\propto 1/q^2$ (with an accuracy $\sim O(1)$) to the $f$ expression and with an accuracy $\sim O(q)$ to the $s$ expression:

\begin{align}
&f=\dfrac{\alpha^2(1+f)\pi^2}{96 T^2 n^2 (1-s)^2}\left(\dfrac{1}{q^2}+O(1)\right);
\label{fgen1}\\
&s\cdot\omega=\dfrac{\alpha^2}{16\pi nT(1-s)} \left(\dfrac{2\pi^3\omega}{12nTq^2}+O(q) \right) \label{sgen1}
\end{align}

\noindent These results were used for derivation (via straightforward algebra) of equations (\ref{sG}) and (\ref{fG}). Constant $G$ defined in (\ref{sG}), was derived directly from expression (\ref{sgen1}), that leads to definition for $G$ in expression (\ref{sG}) by virtue of equations that connect parameter $\alpha$, (\ref{DD}), with parameters $\mu$ and $\lambda$ via expressions (\ref{pars}), (\ref{del}) and (\ref{jacdefs}):

\begin{equation}
G^2\equiv\dfrac{\alpha^2\pi^2}{96q^2n^2T^2}=\dfrac{\mu^2}{24\lambda n^2T^2}.
\label{GG1fin}
\end{equation}  

\subsubsection{Low temperatures limit: $g\gg nT$}
In the limit  $g\gg nT$ we substitute hyperbolic tangents with unity in the numerators in (\ref{fgen}) and (\ref{sgen}), while still expanding trigonometric sine functions in denominators in powers of  small parameter $q$. This leads with an accuracy $\sim O(T/g)$ to the following results:

\begin{align}
&f=\dfrac{\alpha^2(1+f)\pi^2}{2(1-s)^2q^2g^2}\left(1+O\left(\frac{nT}{g}\right)\right);
\label{fgen2}\\
&s\cdot\omega=-\dfrac{\omega\alpha^2\pi^2}{2(1-s)q^2g^2}\left(1+O\left(\frac{nT}{g}\right)\right).\;\label{sgen2}
\end{align}  

\noindent These results were used for derivation (via straightforward algebra) of equations (\ref{sG2})  and (\ref{fG2}). Constant $G_1$ defined in (\ref{sG2}), was derived directly from expression (\ref{sgen2}), that leads to definition for $G_1$ in expression (\ref{sG2}) by virtue of equations that connect parameter $\alpha$, (\ref{DD}), with parameters $\mu$ and $\lambda$ via expressions (\ref{pars}), (\ref{del}) and (\ref{jacdefs}):

\begin{equation}
G^2_1\equiv\dfrac{\alpha^2\pi^2}{2q^2g^2}=\dfrac{2\mu^2}{\lambda g^2}.
\label{GG1fin}
\end{equation}  

\end{document}